\documentclass{ws-ijmpa}
\usepackage[compress]{cite}
\usepackage{graphicx}
\begin{document}

%
%
\title{\boldmath Quantum Entanglement of Fermions- Antifermions Pair Creation modes in Non-commutative Bianchi I Space-time}
\author{M. F. Ghiti}
\address{Laboratoire de Physique Math\'{e}matique et Subatomique,\\ 
Constantine 1 University, Constantine, Algeria \\farouk.ghiti@yahoo.com}
\author{N. Mebarki}
\address{Laboratoire de Physique Math\'{e}matique et Subatomique,\\
Constantine 1 University, Constantine, Algeria \\ nnmebarki@yahoo.fr }
\author{H. Aissaoui\footnote{Corresponding author.}}
\address{Laboratoire de Physique Math\'{e}matique et Subatomique,\\ Constantine 1 University, Constantine, Algeria \\ 
aissaoui\_h@yahoo.com}

\maketitle


\begin{abstract}
The non-commutative Bianchi I curved space-time vierbeins and spin connections are derived. 
Moreover, the corresponding non-commutative Dirac equation as well as its solutions are presented. 
As an application within the quantum field theory approach using Bogoliubov transformations, 
the von Neumann fermion-antifermion pair creation quantum entanglement entropy is studied. 
It is shown that its behaviour is strongly dependent on the value of the non-commutativity 
$\theta$ parameter, $k_\perp$-modes frequencies and the structure of the curved space-time.
 Various discussions of the obtained features are presented.
\end{abstract}

\ccode{PACS numbers: 03.67.-a, 04.62.+v, 03.65.-w, 02.40.Gh, 03.65.Ud}


\section{Introduction}
\label{Int}
During the last few years, non-commutative (N.C.) Seiberg-Witten 
(S.W.) space-time geometry has played an important role in understanding various 
phenomena for example in particle physics and cosmology \cite{1,2,3}. Furthermore, quantum entanglement (Q.E.) 
has been extensively studied in non-relativistic flat-space setups and expanding universes
\cite{4,5,6,7,8,9,10,11,12,13,14,15}. Increasing interest to the emerging 
field of relativistic quantum information and entanglement has attracted 
many people \cite{16a,16,17,18,19,20,21}. Recently, in Refs. \citen{16} and \citen{17}, Q.E. of fermionic and bosonic 
particles in a certain type of Freedman-Robertson-Walker (F.R.W.) universe has been 
shown to have special $k$-modes frequencies and mass dependence. In fact, as it was pointed out in Ref.  \citen{16}, 
the response of Q.E. to the dynamics of the expansion of the universe is affected by the particular choice of 
quantum field theory employed and the geometric structure of space-time. 
Information about the rate and volume of the expansion is codified in the 
frequency and amount of the entangled modes. In our case, as we will see 
later, some of the thermodynamical quantities are better evaluated employing estimation techniques that use Q.E. generated between fermions and antifermions in cosmology scenarios  
which is very sensitive to anisotropies of the universe expansion and 
deformation of the space-time. What we have to know is how much information 
about anisotropies, deformation and thermodynamics is encoded into this 
entanglement created by the dynamics of the Bianchi I space-time. It is worth mentioning that there is no physical 
principle or mathematical formalism which can constrain the order of magnitude of the non-commutativity parameter nor the scale for which the non-commutative model in consideration is relevant. The related energy scale could be as low as a few TeV \cite{22,22a,22b,22c,22d,22e,22f,22g,22h,22i,22j,23}, the same order of magnitude of energies employed in collider experiments (LHC, ILC, etc.), or Planck scale, as it is the case of string or quantum gravity \cite{24,24a,25,26,27,27a}. The N.C. gravity, which we present in this paper, can 
be considered as an effective theory. Thus, the non-commutativity of space-time  can be 
reinterpreted as an extra interaction term on a commutative space-time, and therefore in this way the theory is equivalent to a higher derivative or curvature extension of ordinary gauge gravity\cite{2}. Moreover, 
even if the non-commutativity of space-time  is relevant at the Planck scale, 
the physical phenomena which are induced, like pair creation and entanglement, can 
appear at larger scales. The main goal of this paper is not the derivation of the Dirac equation from the first 
principles and Lagrangian formalism in itself but to study the von Neumann 
Q.E. of fermionic-antifermionic modes created by the dynamics of the N.C. 
Bianchi I universe and understand the new features generated by both, the space-time 
expansion and the non-commutativity $\theta$ parameter. In section \ref{math}, we present the N.C. mathematical 
formalism. In section \ref{Res}, we derive the expression of the bipartite fermion-antifermion 
Q.E. codification of the information as a function of non-commutativity $\theta$ parameter 
and $k_\perp$-modes frequencies as well as the pair creation density. We will see also the role of Q.E. in determining 
some of the thermodynamical properties of this space-time. Finally in section \ref{Conc}, we show the main results and draw our conclusions.
\section{Mathematical Formalism}
\label{math}
The N.C. space-time is characterised by the coordinates operators $
\hat{x}^{\mu}$ $(\mu=\overline{0,3})$ satisfying the following commutation 
relation:
\begin{equation}
\left[ \hat x^{\mu},\hat x^{\nu}\right]=i \,\theta^{\mu\nu}
\end{equation}
where $\theta^{\mu\nu}$ are antisymmetric matrix elements that control the 
non-commutativity of the space-time. The Dirac equation for a massless S.W. spinor 
particle $\hat{\psi}$ in a curved N.C. space-time is shown to take the 
following form (see \ref{appendix A}) \cite{28}:
\begin{equation}\label{Dirac equation}
\left[ \gamma^{f}\left( i\,\partial_{f}+\hat{A}_{f}\right) +\gamma^{f}
\gamma^{5}\hat{B}_{f}\right]\ast\hat{\psi}=0
\end{equation}
where
\begin{eqnarray}
\hat{A}^{f}&=&\Im\left(\hat{e}^{\mu}_{f}\sum_{a=1}^{4}\hat{\omega}^{aa}
_{\mu}\right) 
+\Re\left[ \hat{e}^{\mu}_{d}\left( \hat\omega^{fd}_{\mu}-\hat\omega^{df}
_{\mu}\right) \right] \\
\hat{B}^{f}&=&\Im \left[_{_{_{_{_{_{}}}}}} \left(\hat{e}^{\mu d}\,\hat
\omega^{ab}_{\mu}\right)\right.+\left.
\frac{1}{4}\,\theta^{\rho\sigma}\,\theta^{\alpha\beta}
\left(\partial_{\rho}\partial_{\alpha}\hat{e}^{\mu d}\right) 
\left(\partial_{\sigma}\partial_{\beta}\hat\omega^{ab}_{\mu}\right)\right] 
\varepsilon_{fdab}
\end{eqnarray}
($\Im$ and $\Re$ stand for imaginary and real parts respectively). Here
$\hat{e}^{\mu}_{a}$ and $\hat{\omega}^{fd}_{\mu}$ are the N.C. S.W. 
vierbeins and spin connections respectively (see Ref.  \citen{29}). Their 
general expressions are given in terms of the corresponding commutative 
quantities in \ref{appendix B}. In what follows, the N.C. Dirac matrices $\gamma^{\mu}$ and derivative $
\partial_{\mu}$ in a curved space-time are related to the ones of the Minkowski flat-space through the relations:
\begin{equation}
\gamma^{\mu}=\hat e^{\mu}_{f}\,\gamma^{f}
\end{equation}
and
\begin{equation}
\partial_{\mu}=\hat {e}_{\mu}^{f}\,\partial_{f}
\end{equation}
(Greek and Latin indices are for curved and flat spaces respectively),
where the following vierbeins orthogonality relation holds 
\begin{equation}
\frac{1}{2}\left(\hat e^{a}_{\mu}\ast\hat e^{+\mu}_{b}+\hat e^{\mu}_{b}\ast 
\hat e^{+a}_{\mu}\right)=\delta^{a}_{b}
\end{equation}
In what follows, we consider a Bianchi I universe where the metric has the form:
\begin{equation}\label{metric}
ds^{2}=-dt^{2}+t^{2}\left( dx^{2}+dy^{2}\right)+dz^{2}
\end{equation}
with dimensionless space-time coordinates, (here the time $t$ is 
related to the cosmological parameters of the model). We take the signature convention of 
the space-time to be $(-,+,+,+)$ and for simplicity, make the choice:
\begin{equation}
\theta_{\mu\nu}=\left[
\begin{array}{cccc}
0 & 0 & \theta & 0\\
0 & 0 & 0 & 0\\
-\theta & 0 & 0 & 0\\
0 & 0 & 0 & 0
\end{array}\right]
\end{equation}
Moreover, since we are dealing with massless spinors, it is better to use 
the chiral representation of the Dirac  $\gamma^\mu$ matrices that is:
\begin{eqnarray}
&&\gamma_{0}=\left[
\begin{array}{cc}
\mathbf{0}_{2 \times 2} & \mathbf{1}_{2 \times 2} \\
\mathbf{1}_{2 \times 2} & \mathbf{0}_{2 \times 2} 
\end{array}\right],  
\gamma_{i}=\left[
\begin{array}{cc}
\mathbf{0}_{2\times2} & \sigma_{i 2\times2}\\
-\sigma_{i 2\times2} & \mathbf{0}_{2\times2}
\end{array}\right],
\gamma_{5}=\left[
\begin{array}{cc}
\mathbf{1}_{2\times2} & \mathbf{0}_{2\times2}\\
\mathbf{0}_{2\times2} & -\mathbf{1}_{2\times2}
\end{array}\right]
\end{eqnarray}
where $(i=\overline{1,3})$. Using Maple 16 tensor package, the non-vanishing components of the vierbeins and spin connections have up to 
$O(\theta^{2})$ the following expressions:
\begin{eqnarray}\label{vierbeins}
\hat e^{\tilde{0}}_{0}&=&\hat e^{\tilde{3}}_{3}=1\notag\\
\hat e^{\tilde{1}}_{1}&=&\hat e^{\tilde{2}}_{2}=\frac{1}{t}\Big(1-
\frac{25\,\theta^{2}}{128}\Big)\notag\\
\hat\omega^{13}_{\tilde{2}}&=&-\hat\omega^{12}_{\tilde{3}}=\frac{1}{2}\,\hat
\omega^{31}_{\tilde{2}}
=-\frac{1}{2}\,\hat\omega^{21}_{\tilde{3}}
=-i\,\frac{\theta}{4}\\
\hat\omega^{12}_{\tilde{2}}&=&\hat\omega^{13}_{\tilde{3}}=1+
\frac{5\,\theta^{2}}{128}\notag\\
\hat\omega^{21}_{\tilde{2}}&=&\hat\omega^{31}_{\tilde{3}}=-1+
\frac{7\,\theta^{2}}{64}\notag
\end{eqnarray} 
(the tilde is for a curved space index).
Since the metric presents a space-like singularity at $t=0$, it is 
difficult to define the particle state within the adiabatic approach 
\cite{5,33}. To do so, we first follow a quasi classical approach of Ref.  \citen{34} to 
identify the positive and negative modes frequencies and look 
for the asymptotic behaviour of the solutions at $t\to0$ and $t\to \infty$.
Secondly, we solve the Dirac equation and compare the solutions with the 
above quasi classical limit. Now, in order to find the solutions to the Dirac Eq.(\ref{Dirac equation}) 
in the N.C. Bianchi I space-time, we set:
\begin{equation}
\mathbf{\hat{\psi}}=\left(
\begin{array}{c}
\chi_{1}\\
\chi_{2}\\
\chi_{3}\\
\chi_{4}\
\end{array}\right)
\end{equation}
where
\begin{eqnarray}
\chi_{1}&=&f_{1}(t)\,e^{i \vec{k}\cdot\vec{x}}\notag\\
\chi_{2}&=&f_{2}(t)\,e^{i \vec{k}\cdot\vec{x}}\\
\chi_{3}&=&f_{3}(t)\,e^{-i \vec{k}\cdot\vec{x}}\notag\\
\chi_{4}&=&f_{4}(t)\,e^{-i \vec{k}\cdot\vec{x}}\notag
\end{eqnarray}
the $f_{j}(t)'s$ $(j=\overline{1,4})$ verify the second order differential 
equation:
\begin{equation}
\frac{\partial^{2}f_{j}(t)}{\partial t^{2}}+\frac{A_{j}}{t}\frac{\partial 
f_{j}(t)}{\partial t}+\Big(\frac{C_{j}}{t^{2}}+\frac{D_{j}}{t}-k_{z}^{2}
\Big)f_{j}(t)=0\label{Differential equation}
\end{equation}
with
\begin{eqnarray}
A_{j}&=&1-2\,i\, \Omega^{j}_{2}\notag\\
C_{j}&=&{\Omega^{j}_{1}}^{2}\,k_\perp^{2}+{\Omega^{j}_{3}}^{2}-
{\Omega^{j}_{2}}^{2}\\
D_{j}&=&-i\,k^{j}_{z}-2\,k^{j}_{z}\,\Omega^{j}_{3}\notag
\end{eqnarray}
here
\begin{eqnarray}
k_{\perp}^{2}&=&k_{x}^{2}+k_{y}^{2}\notag\\
k_{z}^{j}&=&(-1)^{j+1}k_{z}\notag\\
\Omega_{1}^{j}&=&\Omega_{1}\\
\Omega_{2}^{j}&=&\Omega_{2}\notag\\
\Omega_{3}^{j}&=&\Omega_{3}(\delta^{j1}-\delta^{j2}-\delta^{j3}+
\delta^{j4})\notag
\end{eqnarray}
and
\begin{eqnarray}
\Omega_{1}&=&-1+\frac{25\,\theta^{2}}{128}\notag\\
\Omega_{2}&=&-2\Big(2-\frac{59\,\theta^{2}}{128}\Big)\\
\Omega_{3}&=&\frac{\theta}{2}\notag
\end{eqnarray}
Now, setting:
\begin{equation}
f_{j}(t)=t^{\alpha_{j}}\,e^{\beta_{j}t}\,h_{j}(t)
\end{equation}
where
\begin{subequations}
\begin{align}
&\alpha_{j}=\frac{1-A_{j}\pm \sqrt{(A_{j}-1)^{2}-4\,C_{j}}}{2}&\\
&\beta_{j}=-\frac{1}{2}& 
\end{align}
\end{subequations}
Eq.(\ref{Differential equation}) can be rewritten as a Kummer's 
differential equation of the form \cite{35}:
\begin{equation}
t\,\frac{d^{2}\,h_{j}(t)}{dt^{2}}+(b_{j}-t)\frac{d\,h_{j}(t)}{dt}-a_{j}\,h_{j}
(t)=0\label{Kummer equation}
\end{equation}
 with
\begin{subequations}
\begin{align}
&a_{j}=-\frac{A_{j}}{2}-\alpha_{j}-D_{j}&\\
&b_{j}=-2\,\alpha_{j}-A_{j}&
\end{align}
\end{subequations}
and the constraint:
\begin{equation}
\vert k_{z}\vert=\frac{1}{2}
\end{equation}
The differential Eq.(\ref{Kummer equation}) has two solutions denoted by 
$M(a_{j},b_{j},z)$ and $U(a_{j},b_{j},z)$ such that:
\begin{subequations}
\begin{eqnarray}
M(a_{j},b_{j},z)&=&\sum_{n=0}^{\infty}\frac{(a_{j})_{n}}{(b_{j})_{n}\,n!}
z^{n}\\
U(a_{j},b_{j},z)&=&\frac{\pi}{\sin(\pi b)}\Big(\frac{M(a_{j},b_{j},z)}
{\Gamma{(1+a_{j}-b_{j})}\,\Gamma{(b_{j})}}
-z^{1-b^{j}}\frac{M(1+a_{j}-b_{j},2-b_{j},z)}{\Gamma{(a_{j})}\,\Gamma{(2-
b_{j})}}\Big)\notag\\
\end{eqnarray}
\end{subequations}
where $(a)_{n}$ is defined by:
\begin{equation}
(a)_{n}=a(a+1)(a+2)....(a+n-1)
\end{equation} 
Note that $(a)_{0}=1$.
Thus, the solution of Eq.(\ref{Kummer equation}) is the following linear 
combination:
\begin{equation}\label{solutions}
f_{j}(t)=C_{1}^{j}\,t^{\alpha_{j}}\,e^{\frac{t}{2}}\,M(a_{j},b_{j},t)+C_{2}^{j}\,
t^{\alpha_{j}}\,e^{\frac{t}{2}}\,U(a_{j},b_{j},t)
\end{equation}
($C_{1}^{j}$ and $C_{2}^{j}$ are constants).
Now, to better understand of the asymptotic behaviour at $t\to0$ 
($''in''$ fields) and $t\to \infty$ ($''out''$ fields) of the solutions Eq.(\ref{solutions}), it is preferable to express the $M(a_{j},b_{j},t)$ and 
$U(a_{j},b_{j},t)$ Kummer functions in terms of Whittaker ones, such that:
\begin{subequations}
\begin{align}
&M\Big(\frac{1}{2}+\mu-\lambda,1+2\mu,z\Big)=e^{\frac{z}{2}}\,z^{-(\frac{1}
{2}+\mu)}\,M_{\lambda,\mu}(z)&\\
&U\Big(\frac{1}{2}+\mu-\lambda,1+2\mu,z\Big)=e^{\frac{z}{2}}\,z^{-(\frac{1}
{2}+\mu)}\,W_{\lambda,\mu}(z)&
\end{align}
\end{subequations}
Note that $M_{\lambda,\mu}(z)$ can be expressed in terms of $W_{\lambda,
\mu}(z)$ as:
\begin{eqnarray}
M_{\lambda,\mu}(z)&=&\frac{\Gamma(2\mu+1)}{\Gamma(\frac{1}{2}+\mu-\lambda)}
e^{-i \pi \lambda}\,W_{-\lambda,\mu}(-z)+\frac{\Gamma(2\mu+1)}{\Gamma(\frac{1}{2}+\mu+\lambda)}e^{-i \pi 
(\lambda-\mu-\frac{1}{2})}\,W_{\lambda,\mu}(z)\notag\\
\end{eqnarray} 
where
\begin{equation}
(W_{\lambda,\mu}(z))^{\star}=W_{-\lambda,\mu}(-z)
\end{equation}  
and
\begin{equation}
(M_{\lambda,\mu}(z))^{\star}=(-1)^{\mu+\frac{1}{2}}\,M_{\lambda,-\mu}(z)
\end{equation}  
Now, to identify the positive and negative modes frequencies in the $''in''$ 
and $''out''$ fields,  we concentrate only on the $f_{1}\equiv f$ function 
(results can be easily extended to $f_{j}, (j=\overline{2,4})$). It is easy 
to show that at $t\to0$, one has:
\begin{subequations}
\begin{align}
&f_{in}^{+}\sim M_{\lambda,\mu}(t)\sim e^{-\frac{t}{2}}\,t^{\mu+\frac{1}{2}}&
\\
&f_{in}^{-}\sim (M_{\lambda,\mu}(t))^{\star}\sim (-1)^{\mu+\frac{1}{2}}\,
M_{\lambda,-\mu}(t)&
\end{align}   
\end{subequations}               
and respectively for $t\to \infty$:
\begin{subequations}
\begin{align}
&f_{out}^{+}\sim W_{\lambda,\mu}(t)\sim e^{-\frac{t}{2}}\,t^{\lambda}&\\
&f_{out}^{-}\sim (W_{\lambda,\mu}(t))^{\star}\sim W_{-\lambda,\mu}(-t)&
\end{align}   
\end{subequations}
where for the both cases we have:
\begin{subequations}
\begin{align}
&\mu=\frac{1}{2}\,(b_{1}-1)&\\
&\lambda=\frac{1}{2}\,(b_{1}-a_{1})&
\end{align}   
\end{subequations} 
A direct consequence of the linear transformation properties of such 
functions is that the Bogoliubov transformations associated with the 
transformation between $''in''$ and $''out''$ solutions take the simple 
form:
\begin{equation}
f_{in}^{\pm}(k_\perp,\theta,t)=\alpha_{k_\perp,\theta}^{\pm}\,f_{out}^{\pm}(k_\perp,\theta,t)+
\beta_{k_\perp,\theta}^{\pm}\,(f_{out}^{\mp}(k_\perp,\theta,t))^{\star}
\end{equation}
where $\alpha_{k_\perp,\theta}^{\pm}$ and $\beta_{k_\perp,\theta}^{\pm}$ are the 
Bogoliubov coefficients. Of course, the curved space spinor solutions of the Dirac equation are 
defined by:
\begin{subequations}
\begin{align}
&\mathbb{U}_{in,out}(\vec{k},\vec{x},t)=f_{in,out}^{-}(k_\perp,\theta,t)\,e^{\imath 
\vec{k}\vec{x}}\,\mathit{U}(0,s)&\\
&\mathbb{V}_{in,out}(\vec{k},\vec{x},t)=(f_{in,out}^{+}(k_\perp,
\theta,t))^{\star}\,e^{-\imath 
\vec{k}\vec{x}}\,\mathit{V}(0,s)&
\end{align}
\end{subequations}      
where $\mathit{U}(0,s)$ and $\mathit{V}(0,s)$ are the ordinary flat space-time spinors, ("$s$" runs over the spin states).
Using the Bogoliubov transformations between the asymptotic terms $''in''$ 
and $''out''$ operators, the field in $''in''$ and $''out''$ regions can 
then be expanded as :
\begin{eqnarray}
\psi_{in,out}&=&\int\frac{d^{3}\vec{k}}{(2\pi)^{\frac{3}{2}}}\sum_{s}
[a_{in,out}(\vec{k},s)\,\mathbb{U}_{in,out}(\vec{k},\vec{x},t)+b_{in,out}^{+}(\vec{k},s)\,\mathbb{V}_{in,out}(\vec{k},\vec{x},t)]
\notag\\
\end{eqnarray}
($a_{in,out}$ and $b_{in,out}^{+}$ are the annihilation and creation 
operators of the fermions and antifermions respectively).  
Due to the form of the Bogoliubov transformation, we can show easily that 
the $''in''$ tensor product of the particle and antiparticle vacuum states 
can be expressed in terms of $''out''$ as:
\begin{equation}
\vert0\rangle_{in}\otimes\vert0\rangle_{in}=\prod_{k_\perp}(A_{0}
\vert0\rangle_{out}\otimes\vert0\rangle_{out}
+A_{1}\vert1_{k_\perp}\rangle_{out}\otimes\vert1_{-k_\perp}\rangle_{out})
\end{equation}                      
where $\vert1_{-k_\perp}\rangle$ (respectively $\vert1_{k_\perp}\rangle$) represents an 
antiparticle (respectively particle) mode with momentum $(-k_\perp)$ 
(respectively $(k_\perp)$). Now, to find the relation between the coefficients 
$A_{0}$ and $A_{1}$ we impose the relation:
\begin{equation}
b_{in}(k_\perp)\vert0\rangle_{in}\otimes\vert0\rangle_{in}=0
\end{equation}                       
yielding to:
\begin{equation}
\alpha_{k_\perp}^{\star}A_{1}\vert1_{-k_\perp}\rangle+\beta_{k_\perp}^{\star}A_{0}\vert1_{-k_\perp}
\rangle=0
\end{equation}   
($\alpha_{k_\perp}$ and $\beta_{k_\perp}$ stand for $\alpha_{k_\perp,\theta}^{\pm}$ and $
\beta_{k_\perp,\theta}^{\pm}$ respectively),                             
and thus, one gets:
\begin{equation}
A_{1}=-\Delta_{k_\perp}^{\star}A_{0}
\end{equation}                                             
where
\begin{equation}
\Delta_{k_\perp}=\frac{\beta_{k_\perp}}{\alpha_{k_\perp}}
\end{equation}                                                      
The normalised vacuum state takes the form:
\begin{equation}
\vert0\rangle_{in}\otimes\vert0\rangle_{in}=\prod\left( 
\frac{\vert0\rangle_{out}\otimes\vert0\rangle_{out}-\Delta_{k_\perp}^{\star}
\vert1_{k_\perp}\rangle_{out}\otimes\vert1_{-k_\perp}\rangle_{out}}{\sqrt{1+\vert
\Delta_{k_\perp}\vert^{2}}}\right) 
\end{equation}                                                               
and it is a pure entangled state of fermion-antifermion modes frequencies with a 
reduced density matrix:
\begin{equation}
\rho_{k_\perp}=\frac{1}{1+\vert\Delta_{k_\perp}\vert^{2}}\Big(\vert0\rangle_{out_{{\,}}out}
\langle0\vert +\vert\Delta_{k_\perp}\vert^{2}\,\vert1_{k_\perp}\rangle_{out_{\,} out}
\langle1_{k_\perp}\vert\Big)
\end{equation}                                                                         
In this case, the entanglement can be quantified for each mode by the von 
Neumann quantum entropy $S(\rho_{k_\perp})$ such that:
\begin{equation}\label{entropy}
S(\rho_{k_\perp})=\log_{2}\left( \frac{1+\vert\Delta_{k_\perp}\vert^{2}}{\vert\Delta_{k_\perp}
\vert^{\frac{2\vert\Delta_{k_\perp}\vert^{2}}{1+\vert\Delta_{k_\perp}\vert^{2}}}}
\right) 
\end{equation}
Eq.(\ref{entropy}) is equivalent to Eq.(21) in Ref. \citen{17}. Note that the quantum entropy can be also expressed in terms of the pair creation density $\hat{n}_{k_\perp}$ as:
\begin{equation}\label{entropy of n}
S(\hat{n}_{k_\perp})=\log_{2}\Bigg(\frac{(\hat{n}_{k_\perp})^{1-2\hat{n}_{k_\perp}}}{(1-
\hat{n}_{k_\perp})^{1-\hat{n}_{k_\perp}}}\Bigg)
\end{equation}
For the N.C. Bianchi I space-time, it is straightforward to show that:
\begin{equation}
\vert\Delta_{k_\perp}\vert^{2}=\Bigg\vert\frac{\Gamma(\frac{1}{2}+\mu+\lambda)}
{\Gamma(\frac{1}{2}+\mu-\lambda)}\Bigg\vert^{2}\vert e^{-i \pi (\lambda-
\mu-\frac{1}{2}})\vert^{2}       
\end{equation}  
It is worth mentioning that the expression of $\vert\Delta_{k_\perp}\vert^{2}$ 
depends strongly on the values of $\mu$ and $\lambda$ (real, complex  and 
pure imaginary) as it is shown in \ref{appendix C}. In the 
case of interest, we distinguish three situations:
\begin{enumerate}
\item[1.] $\theta=0$
\begin{equation}\label{ordinary geometry}
\vert\Delta_{k_\perp}\vert^{2}=\frac{x_{_{}}\sinh(\pi x)}{(1+y)^{2}\,y\,\sinh(\pi y)}
e^{-8\pi k_{\perp}}
\end{equation}
\end{enumerate}
where
\begin{subequations}
\begin{align}
&x=k_{\perp}-\frac{1}{2}&\\
&y=3\,k_{\perp}+\frac{1}{2}&
\end{align}
\end{subequations}
\begin{enumerate}
\item[2.] $\theta>0$
\begin{eqnarray}\label{theta is positif}
\hspace{-1.5cm}\vert\Delta_{k_\perp}\vert^{2}=\left[\frac{[x_{2}\,\Gamma(-x_{2})\,\sin(\pi x_{2})]^{2}
\overset{\infty}{\underset{n=0}{\prod}}\Big(1+\frac{y_{2}^{2}}{(n
+x_{2})^{2}}\Big)}{[x_{1}\,\Gamma(-x_{1})\,\sin(\pi x_{1})]^{2}\overset{\infty}
{\underset{n=0}{\prod}}\Big(1+\frac{y_{1}^{2}}{(n+x_{1})^{2}}\Big)}\right] 
\exp\left(-8\pi(y_{2}+\frac{1}{2})\right)\ \ \ \ \
\end{eqnarray}
\end{enumerate}
\begin{enumerate}
\item[3.] $\theta<0$
\begin{eqnarray}\label{theta is negatif}
\hspace{-1.7cm}\vert\Delta_{k_\perp}\vert^{2}=\left[\frac{\pi^{2}\overset{\infty}
{\underset{n=0}{\prod}}\Big(1+\frac{y_{2}^{2}}{(n+x_{2})^{2}}\Big)}{[x_{1}\,
\Gamma(-x_{1})\,\sin(\pi x_{1})\,\Gamma(x_{2})]^{2}\overset{\infty}
{\underset{n=0}{\prod}}\Big(1+\frac{y_{1}^{2}}{(n+x_{1})^{2}}\Big)}
\right] \exp\left(-8\pi(y_{2}+\frac{1}{2})\right)\ \ \ \ \
\end{eqnarray}
\end{enumerate}
where
\begin{subequations}
\begin{align}
x_{1}=&\frac{\theta}{2}-1\\
x_{2}=&-\frac{\theta}{2}\\
y_{1}=&3\sqrt{\Big(1-\frac{25\,\theta^{2}}{64}\Big)k_{\perp}^{2}+
\frac{\theta^{2}}{4}}+\frac{1}{2}\\
y_{2}=&\sqrt{\Big(1-\frac{25\,\theta^{2}}{64}\Big)k_{\perp}^{2}+
\frac{\theta^{2}}{4}}-\frac{1}{2}
\end{align}
\end{subequations}                                                                                                                                           
Notice the presence of linear terms in the non-commutativity  $\theta$ parameter in 
the expressions of  $x_{1}$ and $x_{2}$. This is due essentially to the 
fact that the S.W. vierbein and spin connection used from Ref.  \citen{29} are 
in general complex (e.g. terms pure imaginary of the spin connection are 
proportional to $\theta$) see Eqs.(\ref{vierbeins}) in our paper together 
with a Dirac wave function $\hat{\psi}$ which is also complex. Thus, we 
expect in addition to second order terms in $\theta$, the  first order ones 
as well coming from the imaginary parts of $\hat{\omega}^{ab}_{\mu}$ and $
\hat{\psi}$. 
\section{Numerical Results and Discussions} 
\label{Res}
\begin{figure}[!th]
        \begin{minipage}{0.49\linewidth}
          \includegraphics[scale=0.37]{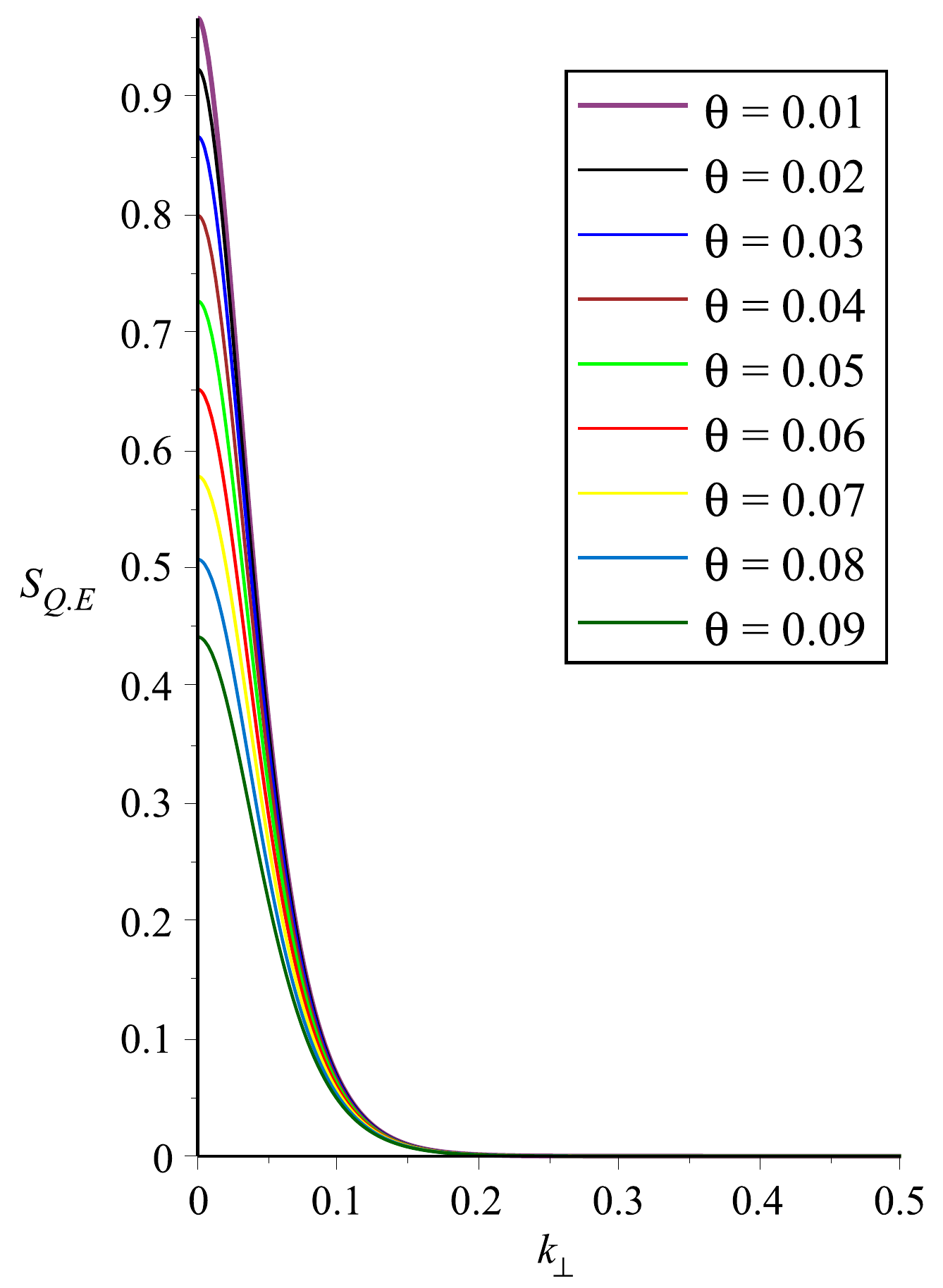} 
          \caption{ $S_{Q.E}$ as a function of the $k_\perp$-modes frequencies for 
various values of $\theta$ parameter in region I}\label{Fig1}
        \end{minipage}
        \hfill
        \begin{minipage}{0.49\linewidth}
          \includegraphics[scale=0.37]{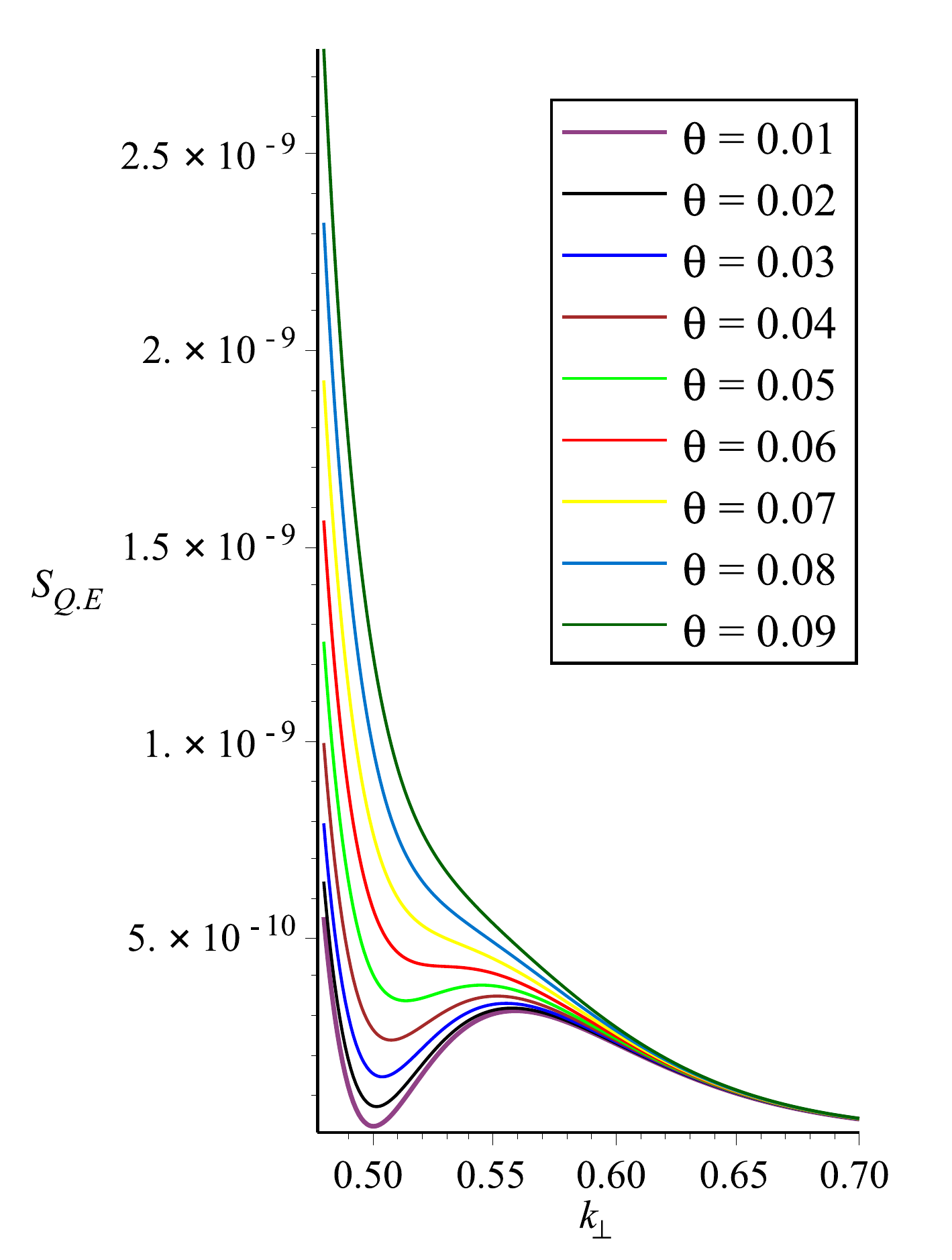}
          \caption{ $S_{Q.E}$ as a function of the $k_\perp$-modes frequencies for various values of $\theta$ parameter in region II}
          \label{Fig2}
        \end{minipage}
      \end{figure}

\begin{figure}[!t]
\begin{minipage}{0.47\linewidth}
\includegraphics[scale=0.3]{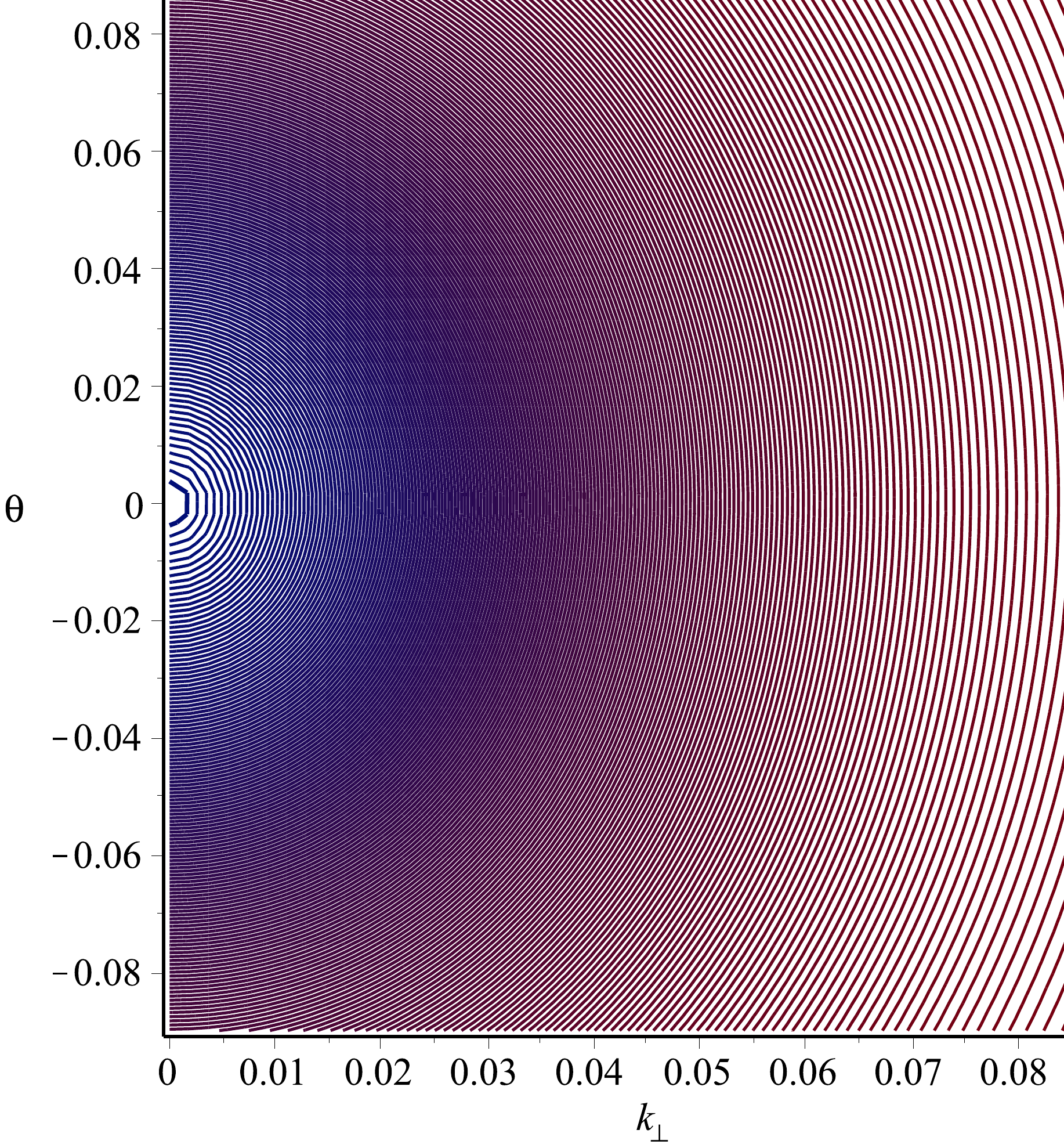} 
\caption{ Contour plots of $S_{Q.E}$ as a function of the $k_\perp$-modes frequencies and $\theta$ parameter}
\label{Fig3}
\end{minipage}
        \hfill
         \begin{minipage}{0.45\linewidth}
\includegraphics[scale=0.47]{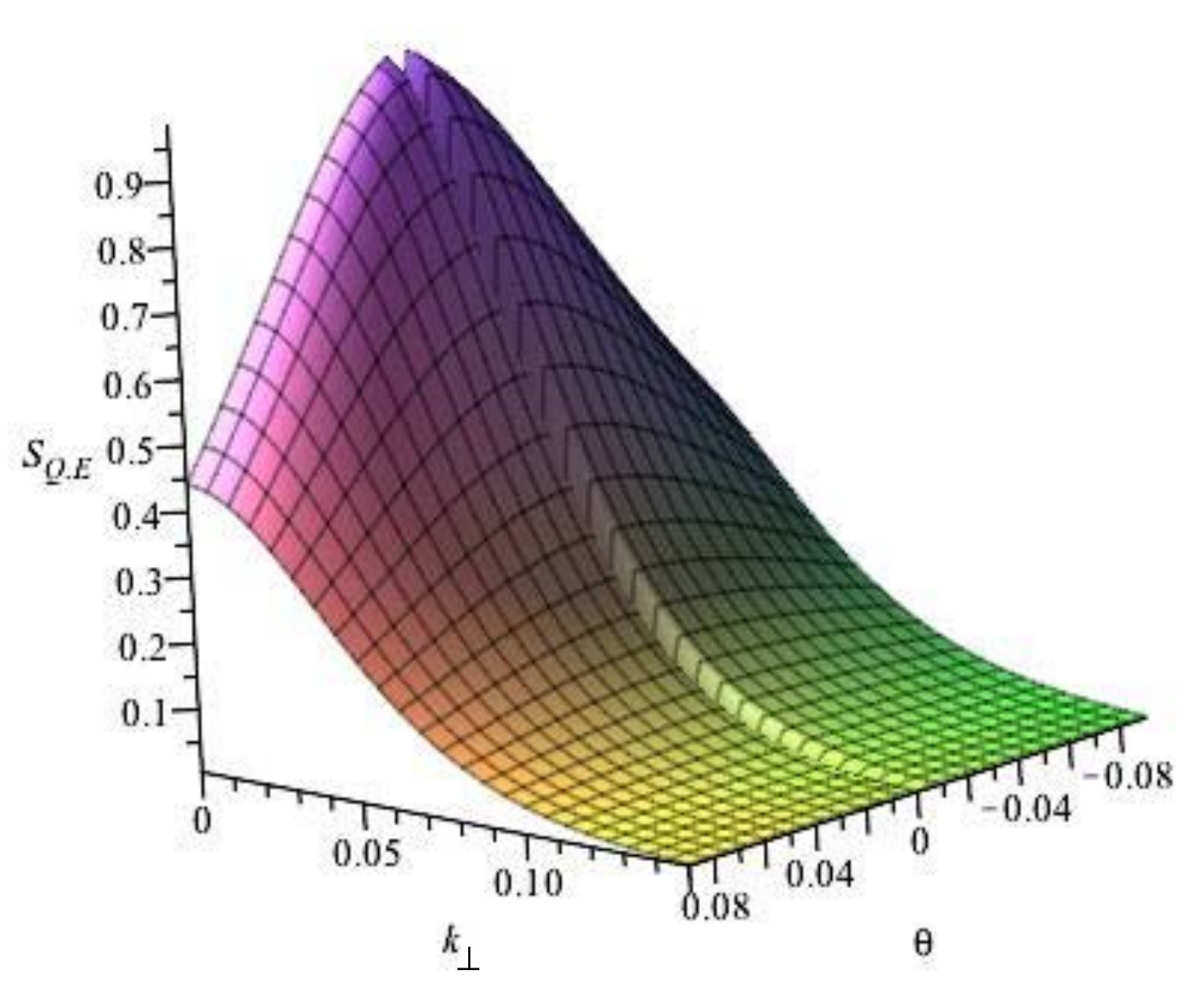} 
\caption{ $S_{Q.E}$ as a function of the $k_\perp$-modes frequencies and $\theta$ parameter}
\label{Fig4}
 \end{minipage}
\end{figure}

By analysing our numerical results concerning the quantum entanglement entropy $S_{Q.E}$ in N.C. Bianchi I 
universe as a function of the $k_\perp$-modes frequencies and non-commutativity $\theta$ parameter, we have 
distinguished two main regions denoted by I and II corresponding to $k_\perp\le\frac{1}{2}$ and $k_\perp>
\frac{1}{2}$ respectively. In region I, $S_{Q.E}$ has a maximal value at $k_\perp=0$. Notice that if $\theta$ is 
relatively small, $S_{Q.E}$ decreases as $k_\perp$ increases until reaching a minimum value at approximately 
$k_\perp\approx \frac{1}{2}$. Then, it increases until reaching a peak (with a very small value) near $k_\perp
\approx0.55$. Finally it decreases again and vanishes at infinity. However, for relatively big values of $\theta$, 
$S_{Q.E}$ becomes a monotonically decreasing function of $k_\perp$ (see FIG.\ref{Fig1} and FIG.\ref{Fig2}). 
Now, for a fixed value of $k_\perp$, we notice that in the region I, $S_{Q.E}$ is a decreasing function of $\theta$. 
However, in the region II, it is an increasing function (see FIG.\ref{Fig1} and FIG.\ref{Fig2}). To be more explicit, 
we have displayed, in the contour plots of FIG.\ref{Fig3} and in the $3$D curve of FIG.\ref{Fig4}, the behaviour 
of $S_{Q.E}$ as a function of $k_\perp$-modes frequencies and non-commutativity $\theta$ parameter. The 
theoretical explanation of such features is that in the region I, one can show that $\Delta_{k_\perp}$ is 
approximately proportional to $e^{-4\pi \theta}$. Therefore, it is clear that if $\theta$ increases, $\Delta_{k_
\perp}$ decreases and since $\frac{\partial{S}}{\partial{\Delta_{k_\perp}}}=-\frac{1}{(1+\Delta_{k_
\perp})^{2}}\,\log_{2}\Delta_{k_\perp}$ is positive for $\Delta_{k_\perp}<1$ (case of our interest), then 
$S_{Q.E}$ decreases (see FIG.\ref{Fig1}). 
The new important numerical result is that the upper bound of the 
fermionic $S_{Q.E}$ (where one has a maximally entangled state) in the physical interval of $k_\perp$ is no 
more $\log_{2}N$ where $N$ is the Hilbert space dimension (in our case $N=2$) corresponding to $k_\perp
\approx \frac{1}{8\pi}\log_{2}(\frac{5}{8})< 0$ but it depends strongly on the non-commutative $\theta$ 
parameter values. It is worth mentioning that in the region I the pair creation density has a non-thermal 
behaviour. In fact, because of the anisotropy of the Bianchi I space-time (which is more complicated than the 
isotropic F.R.W. of Ref.  \citen{16}) and non-commutativity, if a created gas of fermions is observed at a time 
scale much larger (less energy) than the expansion time then the particle number density deviates from the 
quasi equilibrium distribution without a well defined temperature nor a chemical 
potential. Furthermore, if we have an anisotropic space-time, the created particle-antiparticle pair (with the 
same energy) can not reach an equilibrium state in all space directions, except if their energies 
exceed a certain critical value $(k_\perp\approx\frac{1}{2})$ beyond which the 
anisotropic effects become negligible. Thus, for $k_\perp\le\frac{1}{2}$, the particle-antiparticle
pair creation velocity (energy) is less than the expansion velocity in the $x$ and $y$ directions and the 
density of the pair creation is in a non-thermal out-of-equilibrium state. Concerning the privileged value of $k_
\perp$ $(k_\perp=0)$ for which $S_{Q.E}$ is 
maximum, it is related to the characteristic wavelength 
correlated to the underlying space-time structure and non-commutativity 
(deformation). In fact, contrary to the argument (which seems general) given in 
Ref.   \citen{16} using the Pauli exclusion principle which states that it is logical that it is much cheaper to 
excite smaller $k$-modes frequencies in an expanding space-time. Our numerical analysis shows that the 
behaviour of $S_{Q.E}$ as a function of the $k_\perp$-modes frequencies depends not only on the particles 
species (fermions or bosons) but on the space-time structure and deformation as well. It is very important to 
stress on the fact that in the region I, the non-commutativity plays the role of gravity  slowing down the 
expansion and leading to a decrease in the information (quantum entanglement between the fermion-antifermion pair) encoded in $S_{Q.E}$. In the region II, ($k_\perp>\frac{1}{2}$) and contrary to the region I 
($k_\perp\le\frac{1}{2}$), $S_{Q.E}$ is an increasing function of $\theta$ for a fixed value of $k_\perp$. 
Theoretically, the non trivial behaviour of $S_{Q.E}$ can be explained as follows: 
in fact, for relatively small values of $\theta$ and $k_\perp\approx0.5$ (resp. $k_\perp\approx0.55$) we have 
obtained numerically a minimum  (resp. maximum) value of $S_{Q.E}$ . However for relatively large values of $
\theta$, we have checked numerically that  $S_{Q.E}$ 
is a monotonically decreasing function of the $k_\perp$-modes frequencies (see FIG.\ref{Fig2}). Regarding the 
thermal behaviour of the pair creation number density $\hat{n}$ and $S_{Q.E}$, it is important to 
notice that if $\Delta_{k_\perp}\ll1$ or equivalently $k_\perp\gg1$, one can show that $\hat{n}$  
behaves as $e^{-8\pi k_\perp}$ for $\theta=0$ (thermal behaviour) leading to $S_{Q.E}\approx-\Delta_{k_
\perp}\log_{2}\Delta_{k_\perp}$. For $\theta \ne0$, one has $\hat{n}\propto \theta^{2}e^{-8\pi k_\perp}$. It 
is clear that for fixed values of the $k_\perp$-modes frequencies, $\Delta_{k_\perp}$ and $S_{Q.E}$ are 
increasing functions of $\theta$ $(\Delta_{k_\perp}<1)$.
\begin{figure}[!t]
        \begin{minipage}{0.49\linewidth}
          \includegraphics[scale=0.4]{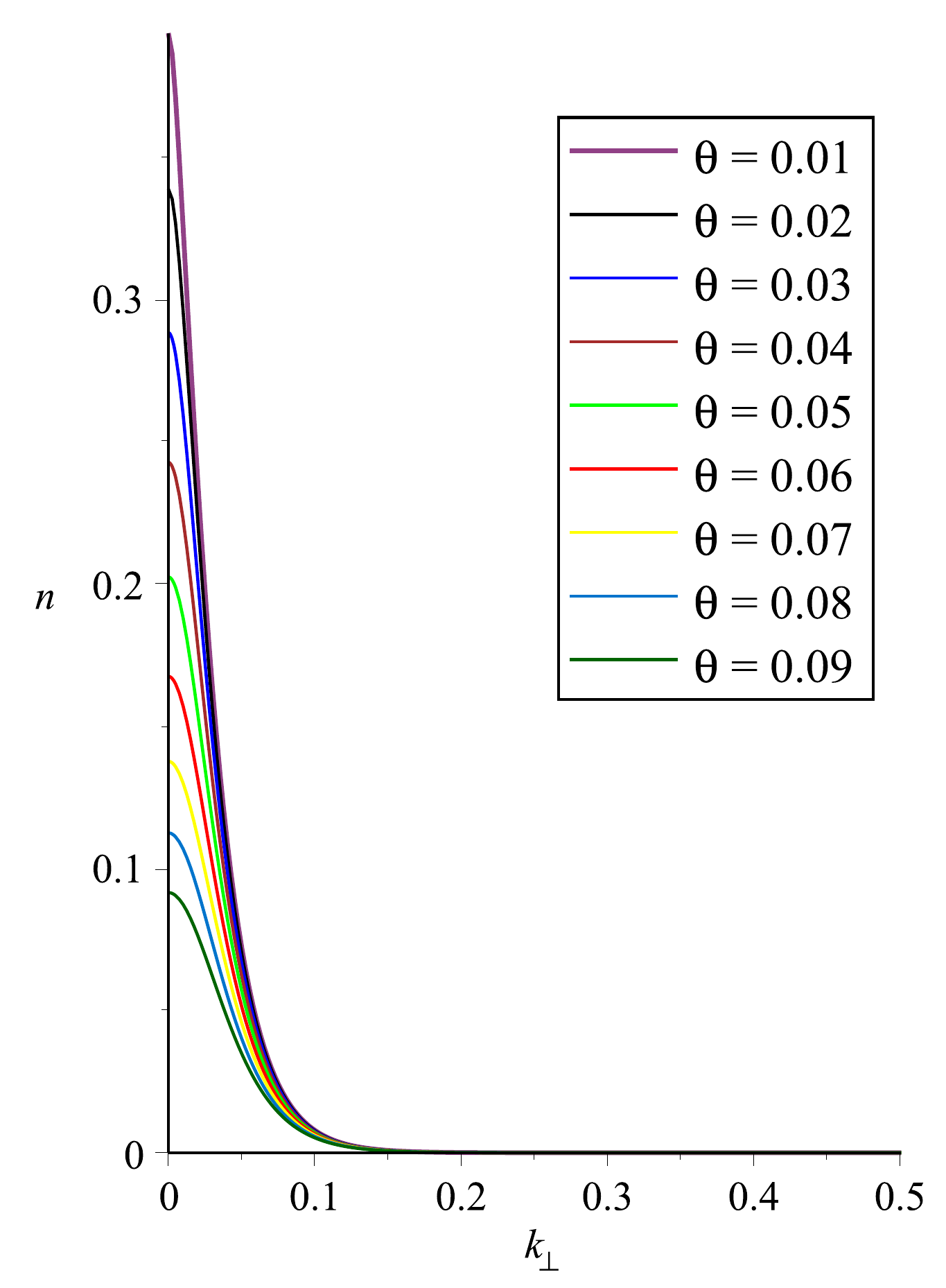} 
          \caption{  $\hat{n}$ as a function of the $k_\perp$-modes frequencies for various values of $\theta$ parameter in 
          region I}\label{Fig5}
        \end{minipage}
        \hfill
        \begin{minipage}{0.49\linewidth}
          \includegraphics[scale=0.4]{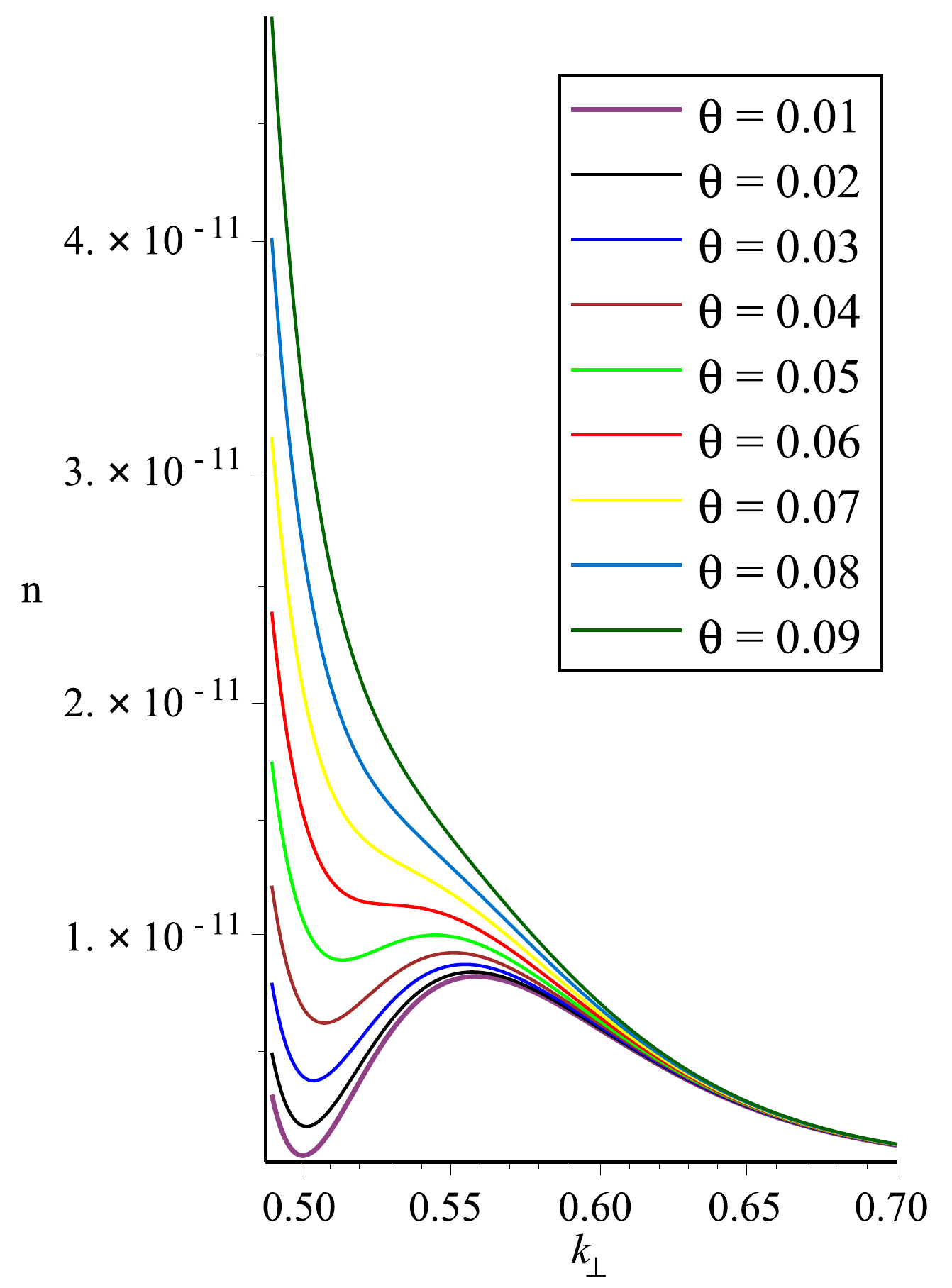}
          \caption{ $\hat{n}$ as a function of the $k_\perp$-modes frequencies for various values of $\theta$ parameter in 
          region II}\label{Fig6}
        \end{minipage}
      \end{figure}
\begin{figure}[!h]
\begin{center}
\includegraphics[scale=0.4]{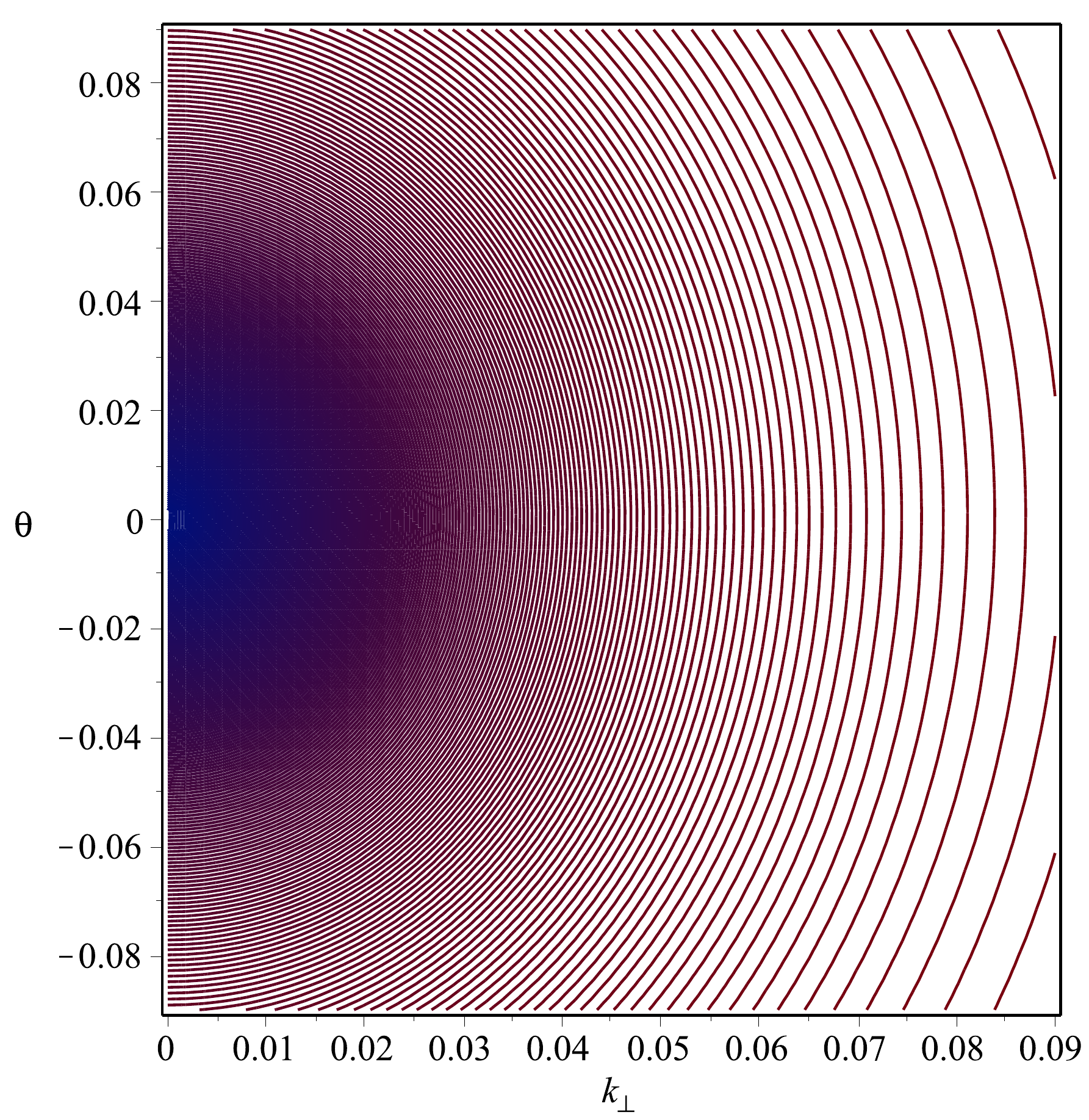} 
\end{center}
\caption{  Contour plots of $\hat{n}$ as a function of the $k_\perp$-modes frequencies and $\theta$ parameter}
\label{Fig7}
\end{figure}
\begin{figure}[!t]
        \begin{minipage}{0.49\linewidth}
          \includegraphics[scale=0.4]{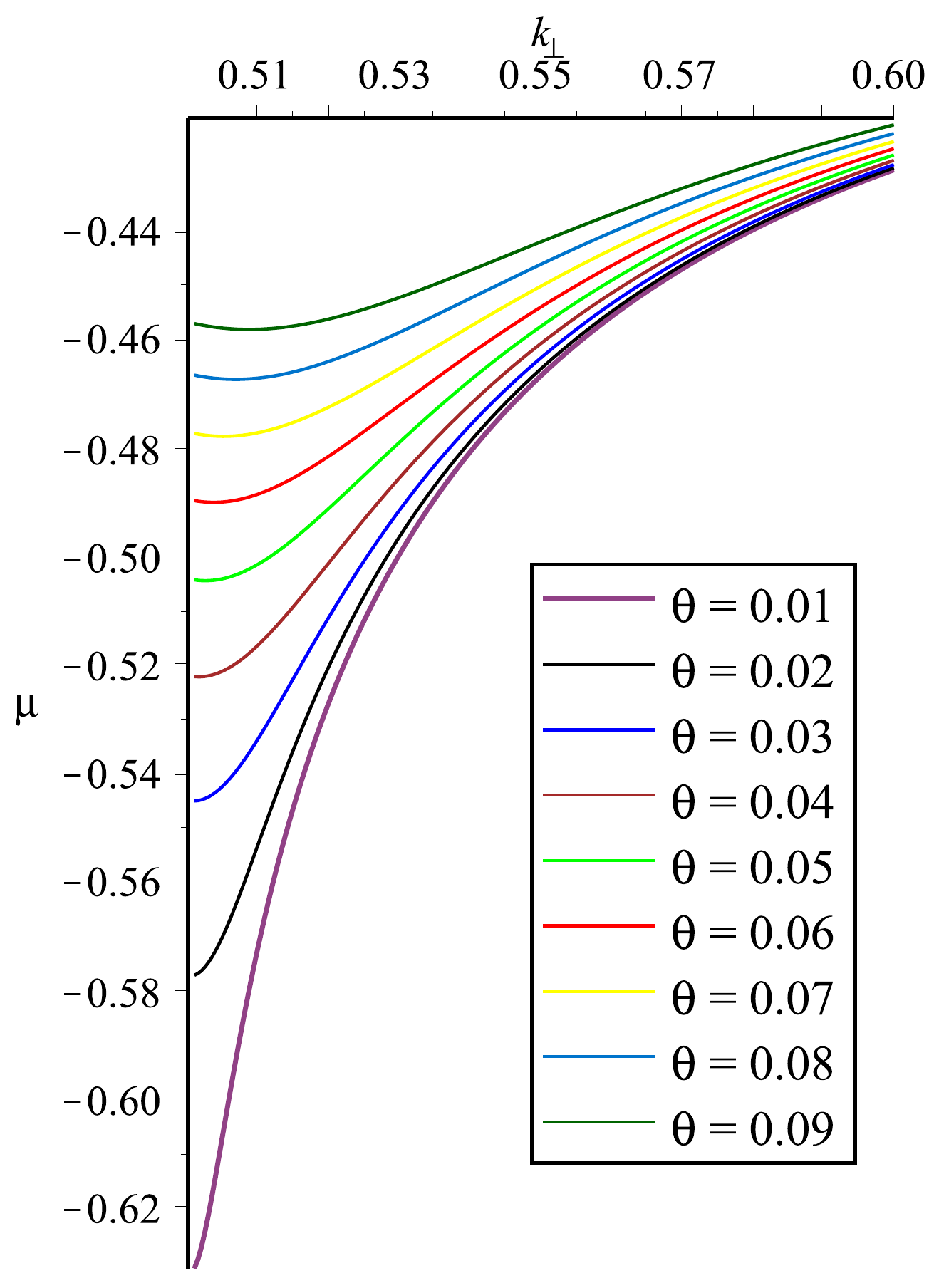} 
          \caption{  $\mu$ as a function of the $k_\perp$-modes frequencies for various 
values of $\theta$ parameter}\label{Fig8}
        \end{minipage}
        \hfill
        \begin{minipage}{0.49\linewidth}
          \includegraphics[scale=0.4]{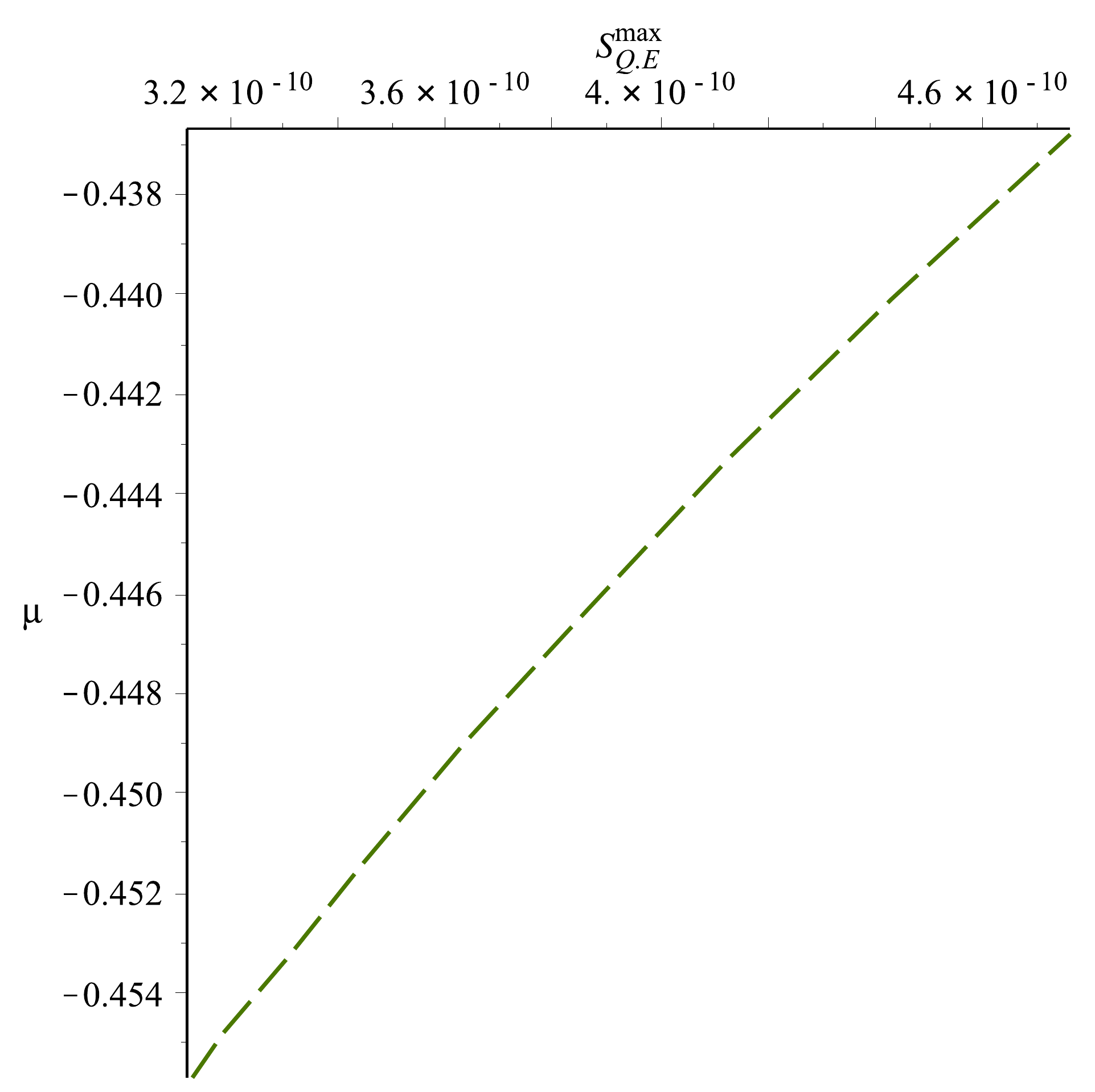}
          \caption{  $\mu$ as a function of the $S_{Q.E}^{max}$}\label{Fig9}
        \end{minipage}
      \end{figure}
      \begin{figure}[!t]
        \begin{minipage}{0.4\linewidth}
          \includegraphics[scale=0.4]{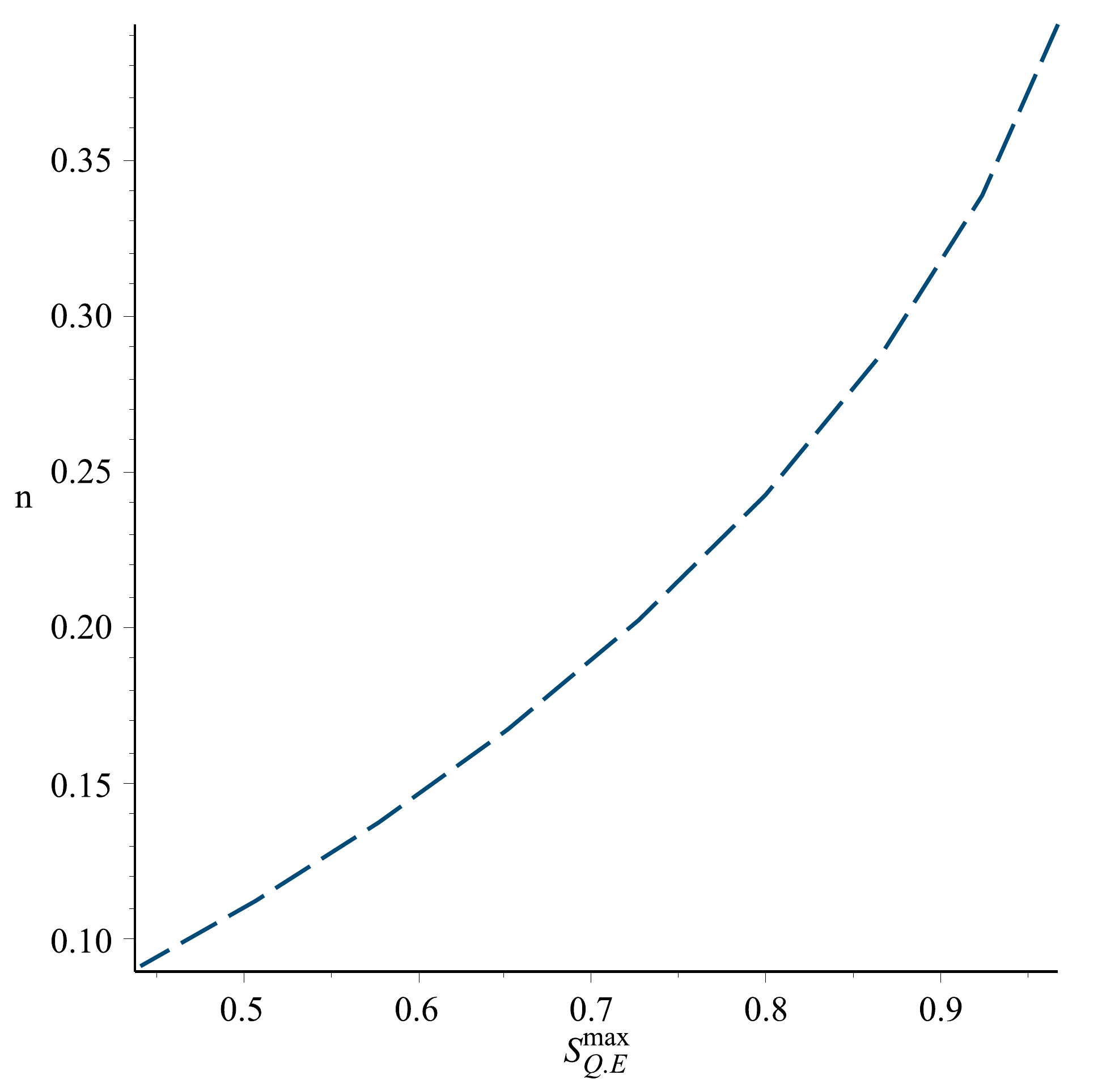} 
          \caption{  $\hat{n}$ as a function of $S_{Q.E}^{max}$}\label{Fig10}
        \end{minipage}
        \hfill
        \begin{minipage}{0.4\linewidth}
          \includegraphics[scale=0.4]{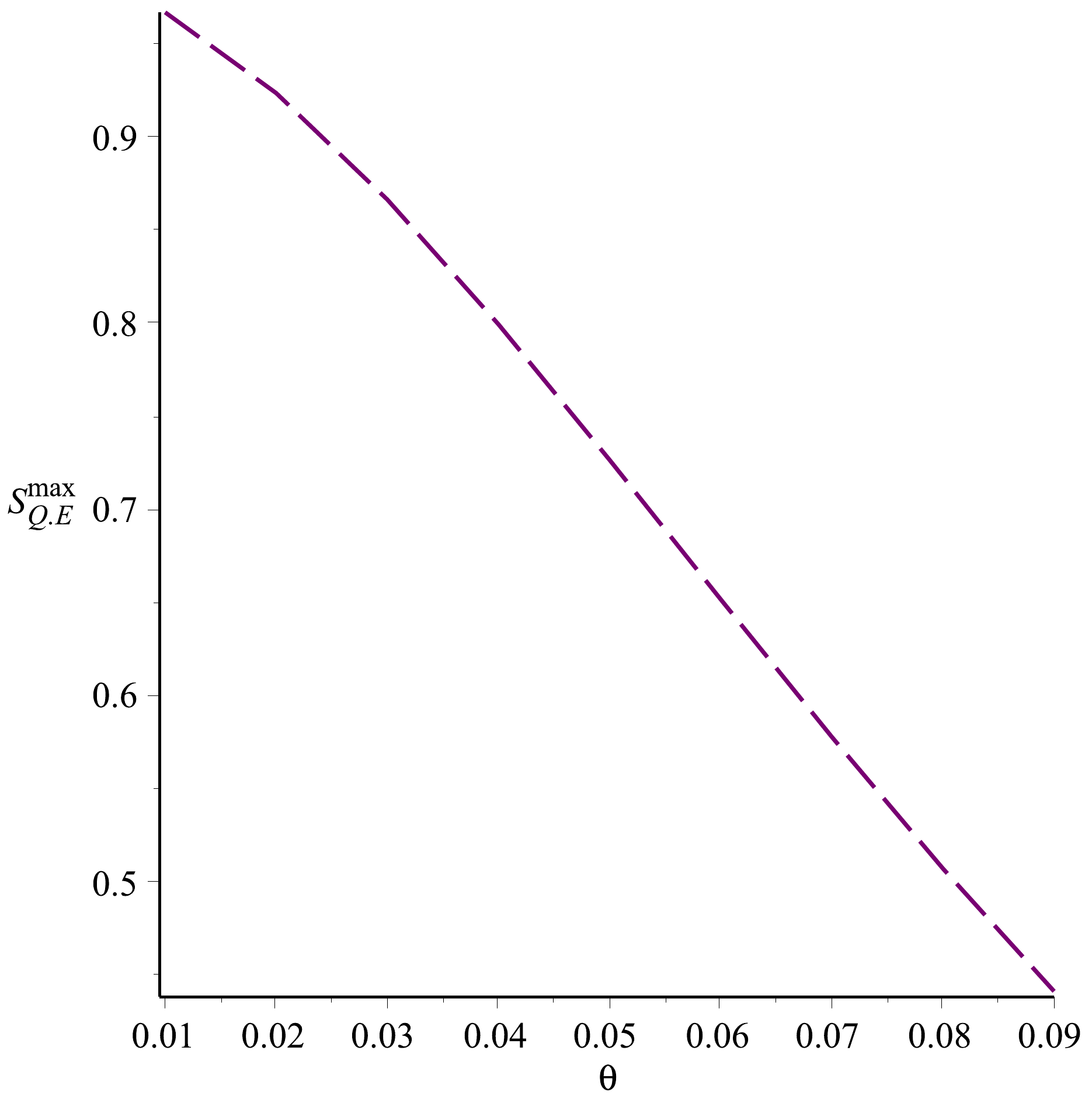}
          \caption{ $S_{Q.E}^{max}$ as a function of $\theta$ parameter}\label{Fig11}
        \end{minipage}
      \end{figure}
This, explains clearly the behaviour shown in FIG.\ref{Fig2}. We remark also that in this region, the non-commutativity 
does not play the role of gravity but rather as a repulsive force (e.g. quintessence, dark 
energy, etc...). In fact, to show the role of the space-time 
non-commutativity $\theta$ parameter and its effect on the quantum 
entanglement, we have noticed that in the region I, $\Delta_{k_\perp}$ is more 
sensitive to the factor $J=e^{-\pi \sqrt{64\,k_\perp^{2}-25\,k_\perp^{2}\,
\theta^{2}+16\,\theta^{2}}}$ (see Eq.(\ref{theta is positif}) and Eq.
(\ref{theta is negatif})). In this case, one has $-25\,k_\perp^{2}\,\theta^{2}+16\,\theta^{2}>0$ 
and therefore, $J$, $\Delta_{k_\perp}$  
and $S_{Q.E}$ become decreasing functions of $\theta$. However, in the region II for large values of $k_\perp$,   
$\Delta_{k_\perp}$ is more affected by the factor $I=\theta^{2}e^{-\pi\sqrt{64\,k_\perp^{2}-25\,k_\perp
^{2}\,\theta^{2}+16\,\theta^{2}}}$. In this case, it is clear that, $I$, $\Delta_{k_\perp}$ and $S_{Q.E}$ 
 are increasing functions of $\theta$ (see FIG.\ref{Fig2}). We conclude that the 
non-commutativity induces two compelling terms with opposite signs:
one is given by the term $+16\,\theta^{2}$ (with a positive sign)
playing the role of gravity or contributing to the matter density through the stress 
energy-momentum tensor and the other is given by the term $-25\,k_\perp^{2}
\,\theta^{2}$ (with a negative sign). Thus, if the first term is bigger 
(case of $k_\perp\le\frac{1}{2}$), the N.C. correction term $-25\,k_\perp^{2}\,
\theta^{2}+16\,\theta^{2} $ will slow down the expansion (like the gravity) 
and the information does not go faster as $\theta$ increases (decrease in 
$S_{Q.E}$). However, for the relatively large values of $k_\perp$, the term 
$-25\,k_\perp^{2}\,\theta^{2}+16\,\theta^{2} $ changes the sign (with respect to the 
first case) and we will have exactly the opposite effect as $\theta$ 
increases. That is $\theta$ will increase the expansion rate and the 
information (quantum entanglement) goes faster (increase in $S_{Q.E}$). It is very important mentioning that since 
$S_{Q.E}$ is an increasing function of the pair creation number density $\hat{n}$ (see Eq.(\ref{entropy of n})) such that:
\begin{equation}
\frac{\partial{S_{Q.E}}}{\partial{\hat{n}}}=-2\log_{2}\hat{n}-\log_{2}(1-\hat{n})+\frac{1-\hat{n}}{\hat{n}}
\end{equation} 
it is always a positive quantity (since $\hat{n}\le1)$. Similar behaviours are obtained for $\hat{n}$. The results are 
summarised in FIG.\ref{Fig4}, FIG.\ref{Fig5}, FIG.\ref{Fig6} and FIG.\ref{Fig7}.
Now, FIG.\ref{Fig8} displays the $k_\perp$-dependence of the chemical potential $\mu$ for various values of $\theta$ (in 
the region II where there is a quasi equilibrium and thermalisation). Notice that $\mu$ is negative and $\vert\mu\vert$ is 
a decreasing function of  $\theta $. The reason is in the thermodynamical equilibrium where $\mu \propto-\frac{\Delta 
S_{Q.E}}{\Delta\hat{n}}$ and since in the region II $\Delta S_{Q.E}$ is an increasing function of $\Delta\hat{n}$, thus the 
ratio $\frac{\Delta S_{Q.E}}{\Delta\hat{n}}$ is positive and therefore $\mu$ is negative.
Moreover, since the increase or decrease of the chemical potential $\mu$ is related to that of $\hat{n}$, we can easily 
explain the $\theta$-dependence of $\mu$ for a fixed value of $k_\perp$. FIG.\ref{Fig9} shows the chemical potential $
\mu$ as a function of the second peak values of  $S_{Q.E}$ (denoted by $S_{Q.E}^{max}$) for various values of $\theta
$. FIG.\ref{Fig10} is the same as FIG.\ref{Fig9} but for the number particle density $\hat{n}$ and the first peak values of 
$S_{Q.E}$ (where $k_\perp\approx 0$). Notice that, if we know the position of the first and second peak, one can get 
easily information about certain thermodynamical quantities (like the chemical potential $\mu$) of the N.C. Bianchi I 
universe. FIG.\ref{Fig11} displays the $\theta$-dependence of the first peak of $S_{Q.E}$. Finally, using the constraint 
$64\,k_\perp^{2}-25\,k_\perp^{2}\,\theta^{2}+16\,\theta^{2}\ge 0$ one can get an upper bound  $\left\vert{\theta}\right\vert\le 
\frac{8}{5}$ which justifies the small numerical values of the non-commutativity $\theta$ parameter considered in our 
paper.  


\section{Conclusion}
\label{Conc}
 Throughout this paper, we have studied the creation of entanglement between massless Dirac fermion-antifermion 
 particle pairs in the framework of N.C. Bianchi I universe characterised by non-commutative components of the 
 vierbeins and spin connections of Eq. (\ref{vierbeins}). In section \ref{math}, we have derived the expressions of these N.C. vierbeins, spin connections and presented  the N.C. Dirac equation and its solutions. Due to the complexity of the N.C. anisotropic Bianchi I  space-time  structure, the behaviours of $S_{Q.E}$ as a function of the $k_\perp$-modes frequencies were not trivial and different from those obtained in Ref.  \citen{16} for the case of an isotropic F.R.W. universe (special solvable case). In fact, at $k=0$, the authors of Ref.  \citen{16} claim that the vanishing of $S_{Q.E}$ (without any rigorous justification) is due to the Pauli exclusion principle and it seems (according to the authors) that it is a general characteristic fermionic feature independent of the space-time structure. According to our numerical results, we notice that the structure and deformation of the space-time as well as the type of the involved particles (fermions or bosons) affect the behaviour of $S_{Q.E}$ and the position of its optimal values as a function of the $k_\perp$-modes frequencies. In Ref.   \citen{16}, the authors have noticed that for 
 massless fermions $S_{Q.E}=0$ (no entanglement) and their results depend on $k=\sqrt{k_{x}^{2}+k_{y}^{2}+k_{z}
 ^{2}}$ (because of the space-time isotropy). Our results show 
 that even with massless particles one can have a non-vanishing entanglement and depend only on $k_{\perp}=
 \sqrt{k_{x}^{2}+k_{y}^{2}}$ (because of Bianchi I space-time anisotropy and the choice of $\theta$). Thus, the
 argument of Ref.  \citen{16} does not hold in general. Lastly, the behaviour of some thermodynamical 
 quantities (like the chemical potential $\mu$) as a function of the optimal $k_\perp$-modes frequencies and 
 deformation of $\theta$ parameter was also discussed. We conclude that a knowledge of the thermodynamical 
 properties of Bianchi I space-time (like chemical potential $\mu$) may give useful information on the entanglement 
 and vice-versa. Thus, to summarise our conclusions, we have shown that:
 \begin{enumerate}
 \item[i.] the behaviour of $S_{Q.E}$ depends not only on the kind of the involved particles (bosons or 
 fermions) during the pair creation process but also on the structure and deformation of the space-time; 
 \end{enumerate}
  \begin{enumerate}
 \item[ii.] because of the space-time deformation the $S_{Q.E}$ does not vanish for massless fermionic particles 
 (contrary to Ref. \citen{16});
  \end{enumerate}
 \begin{enumerate}
 \item[iii.] because of the Bianchi I anisotropy and the choice of the N.C. $\theta$ parameter, our results depend on the  
 $k_\perp$ (transverse) and not on the whole values of $k$;
 \end{enumerate}
 \begin{enumerate}
 \item[iv.] upper new bound of $S_{Q.E}$ depends strongly on the N.C. $\theta$ parameter and not equals to $\log_{2}N$ 
 as it is in Ref. \citen{16};
  \end{enumerate}
  \begin{enumerate}
 \item[v.] for the consistency for our theoretical calculation, we have obtained an upper bound of the N.C. $\theta$ 
 parameter. Of course, the considered small values of $\theta$ in our paper verify this constraint;
 \end{enumerate}
 \begin{enumerate}
 \item[vi.] the non-commutativity of space-time  induces two compelling terms of opposite signs: one plays the role 
 of gravity and contributes to the matter density and the other represents a sort of repulsive force 
 (quintessence, dark energy, etc.). Thus, the information obtained from the quantum entanglement depends on N.C. $\theta$ parameter;
  \end{enumerate}
 \begin{enumerate}
  \item[vii.] if we know the position of the optimal values of the $S_{Q.E}$ we can get information about certain 
 thermodynamical quantities (like the chemical potential  $\mu$) of the N.C. Bianchi I universe and vice-versa.
 \end{enumerate}
 (more studies will be presented in a future publication).                                                                                                                                                                                                         

\section*{Acknowledgments}
We are very grateful to the Algerian Ministry of Higher Education and 
Research and D.G.R.S.D.T. for the financial support. This work is supported 
by C.N.E.P.R.U. under research contract number D00920110030.
\appendix 
\section{The N.C. Dirac Equation}\label{appendix A}
The Dirac equation for a massless fermions in 4-dimensional N.C.S.W. space-time has the following expression:
\begin{equation} \label{mod-dirac eq}
\gamma^{f}\left[ \hat{e}^{\mu}_{f}\,\partial_{\mu}-\frac{i}{8}\left( \hat{e}^{\mu}_{f}\ast\hat{\omega}^{ab}_{\mu}+
\hat{\omega}^{ab}_{\mu}\ast\hat{e}^{\mu}_{f}\right) 
\Sigma_{ab}\right] \ast\hat{\psi}=0
\end{equation}
where $\hat\psi$ is the N.C. Dirac 4-components spinor and:
\begin{equation}
\Sigma_{ab}=\Sigma_{[ab]}+\Sigma_{(ab)}
\end{equation}
with
\begin{eqnarray}
\Sigma_{[ab]}&=&\frac{i}{2}\left[\gamma_{a},\gamma_{b}\right] \\
\Sigma_{(ab)}&=&\frac{1}{2}\left\lbrace \gamma_{a},\gamma_{b}\right\rbrace =\eta_{ab}\,\mathbb{I}_{4\times4}
\end{eqnarray}
using the fact that:
\begin{eqnarray}
\frac{1}{2}\Big[\gamma_{d}\Sigma_{[ab]}+\Sigma_{[ab]}\gamma_{d}\Big]&=&\varepsilon_{fdab}\,\gamma^{f}\gamma^{5}\\
\frac{i}{2}\Big[\gamma_{d}\Sigma_{[ab]}-\Sigma_{[ab]}\gamma_{d}\Big]&=&g_{db}\,\gamma_{a}-g_{da}\,\gamma_{b}\\ 
\notag
\end{eqnarray}
where $\varepsilon_{fdab}$ is the 4-rank totally antisymmetric tensor, one can easily show that the N.C. Dirac equation 
takes the form:
\begin{equation}
\left[ \gamma^{f}\left(i\,\partial_{f}+\hat A_{f}\right) +\gamma^{f}\gamma^{5}\hat B_{f}\right] \ast\hat\psi=0
\end{equation}
where
\begin{eqnarray}
\hat{A}^{f}&=&\Im\left(\hat{e}^{\mu}_{f}\sum_{a=1}^{4}\hat{\omega}^{aa}
_{\mu}\right) 
+\Re\left[ \hat{e}^{\mu}_{d}\left( \hat\omega^{fd}_{\mu}-\hat\omega^{df}
_{\mu}\right) \right] \\
\hat{B}^{f}&=&\Im \left[_{_{_{_{_{_{}}}}}} \left(\hat{e}^{\mu d}\hat\omega^{ab}_{\mu}\right)\right.\left.
\hspace{-0.1cm}+\frac{1}{4}\,\theta^{\rho\sigma}\theta^{\alpha\beta}
\left(\partial_{\rho}\partial_{\alpha}\hat{e}^{\mu d}\right) \left(\partial_{\sigma}\partial_{\beta}\hat\omega^{ab}_{\mu}\right)
\right] 
\varepsilon_{fdab}
\end{eqnarray}
up to $O(\theta^{2})$, 
(The notation $\left[ \  . \ \right]$ stands for the antisymmetric part).\\
The expressions of the S.W. maps $\hat{\omega}_{\mu}$, $
\hat{e}^{\mu}$ and the gauge parameter $\hat{\Lambda}$ (used in our paper are those of Ref.  \citen{29} 
based on the work of Refs. \citen{30,31}) are necessary for the invariance of the action of the pure N.C. gauge 
gravity (which is a part of our total action of Eq. (\ref{total action})) under $\ast$-gauge transformations of Eqs. 
(\ref{eq D-15})-(\ref{eq D-17}). The symmetrisation used in the matter 
field term of the action (\ref{total action}) is due to the $\ast$-product 
ordering ambiguity. As a result and in order to maintain the N.C. gauge 
invariance of the total action, we have 
introduced an N.C. torsion terms (see $2^{nd}$ and $3^{rd}$ terms in the 
action of Eq.(\ref{total action})). Notice that, we can avoid the introduction of the torsion terms and 
preserving the invariance of the total action by modifying the N.C. gauge transformations of the matter field. Of 
course the gauge group is no more $SO(1,3)_\ast$ and therefore the matter fields are not those of S.W. types but 
in the limit $\theta \rightarrow 0$ we recover all ordinary gauge transformations of the Lorentz gauge group 
$SO(1,3)$, (more effects of Moyal $\ast$-product symmetrisation and anti-symmetrisation on the infinitesimal 
gauge transformations and S.W.  maps can be found in Ref.  \citen{32}). It is worth mentioning that even if we 
did  not use a  symmetrised $\ast$-product $\hat{e}^{\mu}_{f}\ast \hat{\omega}^{ab}_{\mu}+\hat{\omega}
^{ab}_{\mu}\ast \hat{e}^{\mu}_{f}$ in Eq.(\ref{mod-dirac eq}), we can not obtain a compact form like for 
example $\gamma^{f}(\hat{e}^{\mu}_{f}\,\partial_{\mu}-\frac{i}{4}\,\hat{e}^{\mu}_{f}\ast \hat{\omega}^{ab}
_{\mu}\,\Sigma_{ab})\ast\hat{\psi}=0$ with $\hat{e}^{\mu}_{f}, \hat{\omega}^{ab}_{\mu}$ and $\hat{\psi} $ 
are S.W. fields from first principles and appropriate Lagrangian density without adding an extra terms like 
torsion, cotorsion etc.

\section{N.C. Mathematical Formalism}\label{appendix B}
The N.C.Vierbeins $\hat e^{a}_{\mu}$ up to $O(\theta^{2})$ are (see Ref.  \citen{29}):
\begin{equation}
\hat e^{a}_{\mu}=e^{a}_{\mu}-i\,\theta^{\nu\rho}\,e^{a}_{\mu\nu\rho}+\theta^{\nu\rho}\,\theta^{\lambda\tau}
\,e^{a}_{\mu\nu\rho\lambda\tau}+O\left( \theta^{3}\right) 
\end{equation}
where
\begin{equation}
e^{a}_{\mu\nu\rho}=\frac{1}{4}\left[ \omega^{ac}_{\nu}\,\partial_{\rho}\,e^{d}_{\mu}+\left( \partial_{\rho}\,\omega^{ac}_{\mu}
+R^{ac}_{\rho\mu}\right) 
e^{d}_{\nu}\right] 
\eta_{cd}
\end{equation}
\begin{eqnarray}
e^{a}_{\mu\nu\rho\lambda\tau}&=&\frac{1}{32}[2\lbrace R_{\tau\nu},R_{\mu\rho}\rbrace^{ab}\,e^{c}_{\lambda}-
\omega^{ab}_{\lambda}(D_{\rho}\,
R^{cd}_{\tau\mu}+\partial_{\rho}\,R^{cd}_{\tau\mu})\,e^{m}_{\nu}\,\eta_{dm} \notag\\ 
&&-\lbrace\omega_{\nu},(D_{\rho}\,R_{\tau\mu}+\partial_{\rho}\,R_{\tau\mu})\rbrace^{ab}\,e^{c}_{\lambda}-\partial_{\tau}
\lbrace\omega_{\nu},(\partial_{\rho}\,\omega_{\mu}+R_{\rho\mu})\rbrace^{ab}\,e^{c}_{\lambda}
\notag\\
&&
-\omega^{ab}_{\lambda}\,\partial_{\tau}(\omega^{cd}_{\nu}\,\partial_{\rho}\,e^{m}_{\mu}+(\partial_{\rho}\,\omega^{cd}_{\mu}
+R^{cd}_{\rho\mu})\,e^{m}_{\nu})\,\eta_{dm}+2\partial_{\nu}\,\omega^{ab}_{\lambda}\,\partial_{\rho}\,\partial_{\tau}\,e^{c}_{\mu}
\notag\\ 
&&
-2\partial_{\rho}
(\partial_{\tau}\,\omega^{ab}_{\mu}+R^{ab}_{\tau\mu})\,\partial_{\nu}\,e^{c}_{\lambda}-\lbrace\omega_{\nu},(\partial_{\rho}
\omega_{\lambda}+R_{\rho\lambda})\rbrace^{ab}\,\partial_{\tau}\,e^{c}_{\mu} \notag\\
&&
-(\partial_{\tau}\,\omega^{ab}_{\mu}+R^{ab}_{\tau\mu})(\omega^{cd}_{\nu}\,\partial_{\rho}\,e^{m}_{\lambda}+(\partial_{\rho}\,
\omega^{cd}_{\lambda}+R^{cd}_{\rho\lambda})\,e^{m}_{\nu}\,\eta_{dm})]\,\eta_{bc} 
\end{eqnarray}
here $R^{ab}_{\mu\nu}$ is the strength field associated with the commutative spin connections $\omega^{ab}_{\mu}$ 
and is defined as:
\begin{equation}
R^{ab}_{\mu\nu}=\partial_{\mu}\,\omega^{ab}_{\nu}-\partial_{\nu}\,\omega^{ab}_{\mu}
+\left( \omega^{ac}_{\mu}\,\omega^{db}_{\nu}-\omega^{ac}_{\nu}\,\omega^{db}_{\mu}\right)\,\eta_{cd}
\end{equation} 
($\eta_{ab}$ is the Minkowski metric).
The N.C. spin connections $\hat\omega^{AB}_{\mu}$ up to $O(\theta^{2})$ are:
\begin{equation}
\hat\omega^{AB}_{\mu}=\omega^{AB}_{\mu}-i\,\theta^{\nu\rho}\,\omega^{AB}_{\mu\nu\rho}
+\theta^{\nu\rho}\,\theta^{\lambda\tau}\,\omega^{AB}_{\mu\nu\rho\lambda\tau}+....
\end{equation}
where
\begin{equation}
\omega^{AB}_{\mu\nu\rho}=\frac{1}{4}\lbrace\omega_{\nu},\partial_{\rho}\,\omega_{\nu}+R_{\rho\mu}\rbrace^{AB}
\end{equation}
\begin{eqnarray}
\omega^{AB}_{\mu\nu\rho\lambda\tau}&=&\frac{1}{32}(-\lbrace\omega_{\lambda},\partial_{\tau}\lbrace\omega_{\nu},
\partial_{\rho}\,\omega_{\mu}
+R_{\rho\mu}\rbrace\rbrace+2\lbrace\omega_{\lambda},\lbrace R_{\tau\nu},R_{\mu\rho}\rbrace\rbrace\notag\\
&&
-\lbrace\omega_{\lambda},\lbrace\omega_{\nu},D_{\rho}\,R_{\tau\mu}+\partial_{\rho}\,R_{\tau\mu}\rbrace\rbrace
-\lbrace\lbrace\omega_{\nu},\partial_{\rho}\,\omega_{\lambda}+R_{\rho\lambda}\rbrace,(\partial_{\tau}\,\omega_{\mu}
+R_{\tau\mu})
\rbrace\notag\\
&&
+2[\partial_{\nu}\,\omega_{\lambda},\partial_{\rho}(\partial_{\tau}\,\omega_{\mu}
+R_{\tau\mu})])^{AB}
\end{eqnarray}
here
\begin{equation}
\lbrace\alpha,\beta\rbrace^{AB}=\alpha^{AC}\,\beta^{B}_{C}+\beta^{AC}\,\alpha^{B}_{C}
\end{equation}
\begin{equation}
[\alpha,\beta]^{AB}=\alpha^{AC}\,\beta^{B}_{C}-\beta^{AC}\,\alpha^{B}_{C}
\end{equation}
and
\begin{equation}
D_{\mu}\,R^{AB}_{\rho\sigma}=\partial_{\mu}\,R^{AB}_{\rho\sigma}+(\omega^{AC}_{\mu}+
R^{DB}_{\rho\sigma}+\omega^{BC}_{\mu}\,R^{DA}_{\rho\sigma})\,\eta_{CD}
\end{equation}

\section{Derivation of N.C. Dirac Equation From a Least Action}

\label{appendix D}

To derive the N.C. Dirac equation given by Eq.(\ref{Dirac equation}) from 
the least action principle,  we consider the following N.C. action:
\begin{eqnarray}\label{total action}
S&=&\int d^{4}x\,\sqrt{-\hat{g}}\ast\left\lbrace \left[\frac{i}{2}\left( 
\bar{\hat{\psi}}\ast\gamma^{\mu}\ast\tilde{D}_{\mu}\right)\ast\hat{\psi}
+c.c\right]\right.+\frac{i}{4}\bar{\hat{\psi}}\ast\hat{K}_{\mu\nu\rho}\ast\hat{X}^{\mu\nu
\rho}\ast{\hat{\psi}}\notag\\ 
&&\left.-\hat{T}^{\mu\nu\rho}\ast\hat{K}_{\mu\nu\rho}\right.\left.+\mathcal{L}_{g}\frac{}{}\right\rbrace 
\end{eqnarray} 
where $\hat{K}_{\mu\nu\rho}$  and  $\hat{T}^{\mu\nu\rho}$  are the N.C. 
cotorsion and modified torsion tensors respectively, and $\hat{g}$ is the 
determinant of the N.C. metric. \\
We set:
\begin{equation}
\frac{i}{4}\,\hat{K}_{\mu\nu\rho}\,\hat{X}^{\mu\nu\rho}=\hat{K}
\end{equation} 
such that
\begin{eqnarray}\label{cotorsion}
\hat{K}&=&\sqrt{-\hat{g}}\ast\left\lbrace\frac{i}{2}\,\gamma^{a}\,\hat{Y}_{a}+
\frac{i}{2}\,\hat{e}^{\mu}_{b}\ast\hat{Y}_{a}\ast\hat{e}^{b}_{\mu}\,\gamma^{a} 
\right\rbrace-\frac{i}{2}\,\partial{\mu}\left( \sqrt{-\hat{g}}\ast
\hat{e}^{\mu}_{a}\right)\gamma^{a}
\end{eqnarray}
and 
\begin{equation}
\hat{Y}_{f}=\frac{i}{8}\left\lbrace \hat{e}^{\mu}_{f},\hat{\omega}^{ab}
_{\mu}\right\rbrace_{\ast}\Sigma_{ab}
\end{equation}
where
\begin{equation}
\left\lbrace a,b \right\rbrace_{\ast}=a\ast b+b\ast a
\end{equation}
here the N.C. covariant derivative $\tilde{D}_{\mu}$ is defined such that  
\begin{equation}
 \hat{e}^{\mu}_{a}\ast\tilde{D}_{\mu}\ast\hat{\psi} = \left(\hat{e}^{\mu}
 _{a}\ast\partial_{\mu}+\hat{Y}_{a} \right)\ast\hat{\psi}
\end{equation}
one can show easily that the variational principle of the action with 
respect to $\hat{\psi}$ gives our N.C. Dirac equation (Eq.\ref{Dirac equation}).
For a comparative illustration, one can get from Eq.(\ref{cotorsion}) in 
the commutative case the following expression:  
\begin{equation}
\hat{K}\to K=\frac{i}{4}\,\omega_{abc} \left\lbrace  \gamma^{[a}\gamma^{b}
\gamma^{c]}-\gamma^{a}\gamma^{[b}\gamma^{c]}   \right\rbrace+\frac{i}{2}\,
\gamma_b{}^b{}_{a}\,\gamma^{a}
\end{equation}
where
\begin{equation}
\gamma_b{}^b{}_{a}=C_{ab}{}^b-C_{ba}{}^b=2C_{ab}{}^b
\end{equation}
and
\begin{equation}
C_{ab}{}^b=\frac{1}{2}\,e^{\mu}{}_{a}\left(  e^{\nu}{}_{b}\,\partial_{\mu}\,
e_{\nu}{}^{b}\right)-\frac{1}{2}\,e^{\mu}{}_{a}\,e^{\nu}{}_{b}\,\partial_{\nu}\,
e_{\mu}{}^{b}
\end{equation}
$\left[ \  . \ \right]$ stands for antisymmetrisation with respect to the 
indices. Moreover, in this case one can show also that:
\begin{equation}
\partial_{\nu}\left(\sqrt{-g}\,e^{\mu}{}_{a}\right)=\sqrt{-g}\,\gamma_b{}^b{}
_{a}
\end{equation}
the cotorsion and the modified torsion tensors are given by:
\begin{equation}
K_{\alpha\beta\mu}=-Q_{\alpha\beta\mu}-Q_{\mu\alpha\beta}+Q_{\beta\mu
\alpha}
\end{equation}
and
\begin{equation}
T_{\mu\nu}{}^{\alpha}=Q_{\mu\nu}{}^{\alpha}+\delta_{\mu}{}^{\alpha}\,Q_{\nu}-
\delta^{\alpha}{}_{\nu}\,Q_{\nu}
\end{equation}
respectively. Here $Q_{\alpha\beta\mu}$ is the torsion tensor and:
\begin{equation}
Q_{\alpha}=Q_{\alpha\nu}{}^{\nu}
\end{equation}
 the N.C. modified torsion tensor $\hat{T}_{\mu\nu}{}^{\alpha}$ has a 
 similar form as in the commutative case:
\begin{equation}
\hat{T}_{\mu\nu}{}^{\alpha}=\hat{Q}_{\mu\nu}{}^{\alpha}+\delta_{\mu}{}
^{\alpha}\,\hat{Q}_{\nu}-\delta^{\alpha}{}_{\nu}\,\hat{Q}_{\mu}
\end{equation}
Concerning the invariance of the N.C. Lagrangian density with respect to 
the N.C. local Lorentz transformations (see Eqs.(\ref{eq 
D-15})-(\ref{eq D-17})), one can choose:
$\hat\delta_{\hat\Lambda}\hat K_{\mu\nu\rho}$  and, or  $\hat{T}^{\mu\nu\rho}= \hat{g}_{\rho\alpha}
\ast\hat{T}_{\mu\nu}{}^{\alpha}$(not all the N.C. components of $\hat K_{\mu\nu\rho}$  
are contributing to $\hat K$) such that:
\begin{eqnarray}
\hat{\delta}_{\hat{\Lambda}}\,\hat{\psi}&=&\hat{\Lambda}\ast\hat{\psi}
\label{eq D-15}\\ 
\hat{\delta}_{\hat{\Lambda}}\,\hat{e}^{\mu}&=&\hat{\Lambda}\ast\hat{e}^{\mu}-
\hat{e}^{\mu}\ast\hat{\Lambda}=\left[\hat{\Lambda},\hat{e}^{\mu}
\right]_{\ast}\label{eq D-16}\\ 
\hat{\delta}_{\hat{\Lambda}}\,\hat{\omega}_{\mu}&=&-\partial_{\mu}
\hat{\Lambda}+\left[\hat{\Lambda},\hat{\omega}\right]_{\ast}\label{eq D-17}
\\ \notag
\end{eqnarray}
and
\begin{equation}
\hat{\delta}_{\hat{\Lambda}}\left[ \bar{\hat{\psi}}\,\hat{K}\,\hat{\psi}\right] 
+\hat{\delta}_{\hat{\Lambda}}\left( \hat{T}^{\mu\nu\rho}\,\hat{K}_{\mu\nu
\rho}\right) +\hat{\delta}_{\hat{\Lambda}}\,\hat{\mathcal{L}_{0}}=0
\end{equation}
where $\left[A,B\right]_{\ast}$ stands for $A\ast B-B\ast A$, and

\begin{eqnarray}
\hat{e}^{\mu}&=&\gamma^{\mu}=\hat{e}^{\mu}{}_{a}\,\gamma^{a}\\
\hat{\omega}_{\mu}&=&\hat{\omega}^{ab}{}_{\mu}\,\Sigma_{ab}\\
{\hat{\Lambda}}&=&{\hat{\Lambda}}^{ab}\,\Sigma_{ab}\\ \notag
\end{eqnarray}
Here $\hat{\delta}_{\hat{\Lambda}}\hat{\mathcal{L}_{0}}$ is the non 
vanishing part of $\hat{\delta}_{\hat{\Lambda}}\left[\frac{i}{2}\,
\bar{\hat{\psi}}\ast\gamma^{\nu}\ast\tilde{D}_{\mu}\ast\hat{\psi}+c.c
\right]$ when we use the N.C. gauge transformations of Eqs.(\ref{eq 
D-15})-(\ref{eq D-17}).  Of course $\hat{\delta}_{\hat{\Lambda}}
\hat{\mathcal{L}_{g}}=0 $ (see Ref.  \citen{30})

This is a simple justification for the derivation of our N.C. Dirac 
equation (Eq.\ref {Dirac equation}) with S.W. fields and proof of 
invariance  of our N.C. Lagrangian density with respect to the N.C. local  
Lorentz transformations of Eqs.(\ref{eq D-15})-(\ref{eq D-17}).

\section{Determination of $\vert\Gamma(x+i y)\vert^{2}$}\label{appendix C}

\begin{enumerate}
\item If $x=0$
\begin{equation}
\vert\Gamma(i\,y)\vert^{2}=\frac{\pi}{y\,\sinh(\pi y)}
\end{equation}
\item If $y=0$
\begin{enumerate}
\item $x>0$
\begin{equation}
\vert\Gamma(x)\vert^{2}=[(x-1)!]^{2}
\end{equation}
\item $x<0$
\begin{equation}
\vert\Gamma(x)\vert^{2}=\frac{\pi^{2}}{[-x\,\Gamma(-x)\,\sin(\pi x)]^{2}}
\end{equation}
\end{enumerate}
\item $x\ne0$, $y\ne 0$
\begin{equation}
\vert\Gamma(i\,y)\vert^{2}=\prod_{n=0}^{\infty}\Bigg[\Big(1+\frac{y^{2}}{(n
+x)^{2}}\Big)\Bigg]^{-1}
\end{equation}
\item If $\frac{y^{2}}{x^{2}}<1$ $(x\ne0)$, then $\vert\Gamma(x+i\,y)\vert$ can be approximated by:
\begin{equation}
\vert\Gamma(x+i\,y)\vert \approx \vert\Gamma(x)\vert \Big[1-\frac{y^{2}}{2}
\Phi(1,2,x)\Big]
\end{equation} 
where $\Phi(1,2,x)$ is the Herwitz zeta function such that:
\begin{equation}
\Phi(1,2,x)=\sum_{n=0}^{\infty}\frac{1}{(x+n)^{2}}
\end{equation}
\end{enumerate}


\end{document}